\documentclass[aps,prl,twocolumn,superscriptaddress]{revtex4-2}  
\usepackage{amsfonts}
\usepackage{bbding}
\usepackage{amssymb}
\usepackage[title]{appendix}
\usepackage{color}
\usepackage{amsmath}
\usepackage{graphicx}
\usepackage{subfigure}
\usepackage{txfonts}
\usepackage{bm}
\usepackage[normalem]{ulem}
\usepackage{natbib}
\usepackage{calrsfs}
\usepackage{tikz}
\usetikzlibrary{graphs, positioning, quotes} 
\usetikzlibrary{matrix}
\usepackage[colorlinks,linkcolor=blue,citecolor=blue,urlcolor=blue]{hyperref}
\usepackage{orcidlink}
\preprint{APS/123-QED}

\newcommand{\resetappendix}{%
    \appendix
    \renewcommand{\thesection}{A\arabic{section}}
    \renewcommand{\theequation}{A\arabic{equation}}
    \renewcommand{\thefigure}{A\arabic{figure}}
    \renewcommand{\thetable}{A\arabic{table}}
    \setcounter{section}{0}
    \setcounter{equation}{0}
    \setcounter{figure}{0}
    \setcounter{table}{0}
}

\DeclareMathAlphabet{\pazocal}{OMS}{zplm}{m}{n}

\newcommand{\Hs}{\pazocal{H}}
\newcommand{\Bs}{\pazocal{B}}
\newcommand{\Ds}{\pazocal{D}}

\newcommand{\Ss}{\pazocal{S}}

\newcommand{\blueit}[1]{\textcolor{blue}{\textit{#1}}}

\begin{document}

\title{Synchronization Mpemba Effect}

\author{Yan Li\orcidlink{0009-0007-7218-133X}}
\affiliation{School of Physics and Information Technology, Shaanxi Normal University, Xi’an, 710062, P. R. China}

\author{Wenlin Li\orcidlink{0000-0003-2032-0530}}
\affiliation{College of Sciences, Northeastern University, Shenyang 110819, China}
\affiliation{Research Center for Quantum Physics, Northeastern University, Shenyang 110819, China}

\author{Xingli Li\orcidlink{0000-0003-2339-6557}}
\email{lixingli0130@gmail.com}
\affiliation{Department of Physics, The Chinese University of Hong Kong, Shatin, New Territories, Hong Kong, China}
\affiliation{The Chinese University of Hong Kong Shenzhen Research Institute, Shenzhen 518057, China}
\affiliation{Lanzhou Center for Theoretical Physics, Key Laboratory of Theoretical Physics of Gansu Province, Key Laboratory of Quantum Theory and Applications of MoE, Gansu Provincial Research Center for Basic Disciplines of Quantum Physics, Lanzhou University, Lanzhou 730000, China}

\begin{abstract}
We introduce the synchronization Mpemba effect, in which a less synchronized initial state overtakes a more synchronized one, and demonstrate it in dissipatively coupled quantum van der Pol oscillators.
Within a graph-Laplacian framework, we show that for two oscillators, the mechanism is a competition between a slowly decaying collective background and a fast transverse transient, equivalently an exchange between relative-mode displacement and phase-fluctuation errors.
The same decomposition extends to networks: the synchronization error splits into transverse Laplacian modes, yielding closed-form crossing times from eigenvalue separations and initial mode weights, and a sufficient screening criterion.
With only these eigenvalues entering, the rule is topology independent: crossing times of different networks collapse onto a single spectral-gap line.
Thus, quantum many-body dynamics can be designed and controlled at low dimension without constructing the exponentially large Liouvillian, opening a route to quantum synchronization as a resource.
\end{abstract}
\date{\today}

\maketitle

\blueit{Introduction.--}
Synchronization occurs when self-sustained oscillators acquire a stable relative phase~\cite{Pikovsky2001}.
Dissipatively coupled quantum van der Pol (vdP) oscillators provide a minimal setting, in which local gain and loss sustain noisy limit cycles while the coupling suppresses relative-phase fluctuations~\cite{Lee2013a,Lee2014a,Walter2014a,Walter2015,Li2020,Schmolke2022,Lorch2017,Li2022a,Solanki2026,Eshaqi-Sani2020a}; quantum synchronization has recently been realized experimentally~\cite{Liu2026n,Li2025d,Laskar2020,Krithika2022,Zhang2023j}.
Synchronized states are a resource for quantum metrology and networked devices, and the shorter the transient that prepares them, the sooner they become available within the finite coherence and control times of an experiment~\cite{Komar2014,Aifer2024,Vaidya2025,Roulet2018}.
For a fixed dissipative generator that transient is set by the relaxation modes of the same open-system dynamics, and shortening it by redesigning the coupling changes the physical resources.
Can a synchronized state instead be prepared faster by choosing a more favorable initial state, with the dissipative dynamics untouched?

The Mpemba effect suggests just such a route: under identical dynamics, a state that starts farther from the steady state can overtake one that starts closer during relaxation~\cite{Mpemba1969,Klich2019,Teza2026}.
In closed quantum systems, the analogue appears in the restoration of a symmetry that the initial state breaks~\cite{Ares2023,Rylands2024,Ares2025,Bhore2025,Joshi2024a,Turkeshi2025,Liu2024,Chang2026}.
In open systems, the effect is encoded in the Liouvillian spectrum: when the initial state has vanishing projection onto the slowest-decaying mode, faster modes control the asymptotic relaxation and the approach to stationarity is exponentially accelerated~\cite{Carollo2021,Li2025r,Medina2025,Zhang2025,AharonyShapira2024,Strachan2025,Bao2025,Moroder2024,Nava2024,VanVu2025a,Zhang2026o,Zhan2026a}.
We therefore define the synchronization Mpemba effect (SMpE), in which an initially less synchronized state overtakes a more synchronized one while both approach the same phase-locked limit cycle. Applying this criterion, however, requires the Liouvillian modes: for many-body networks, the Liouvillian dimension grows exponentially with the number of oscillators, and direct diagonalization is prohibitive.

In this Letter, we overcome this obstacle with the graph Laplacian, whose dimension is exponentially smaller than that of the Liouvillian.
Our main findings are threefold.
(i) For two oscillators, competition between a slowly decaying collective background and a fast transverse transient produces the SMpE, yielding a closed-form crossing time and a sufficient occurrence criterion.
(ii) In networks, the synchronization error resolves into transverse Laplacian modes, so that crossing times and a screening criterion follow from the graph spectrum and the initial mode weights.
(iii) The network topology enters only through the Laplacian spectrum: the crossing times of different networks collapse onto a common scaling, enabling spectrum-based design.

\blueit{Model.--}
We consider $N$ dissipatively coupled quantum vdP oscillators, whose dynamics is governed by the Lindblad master equation
\begin{equation}
\dot{\rho} = -i[H, \rho] + \sum_{k=1}^{N}\left(\gamma_{\uparrow}\Ds[a^{\dagger}_{k}]\rho+\gamma_{\downarrow}\Ds[a_{k}^{2}]\rho\right)+\sum_{k<j}\kappa_{kj}\Ds[a_{k}-a_{j}]\rho,
\label{Eq:ME}
\end{equation}
where $H=\omega\sum_{k}a_k^\dagger a_k$ is the Hamiltonian with oscillator frequency $\omega$ ($\hbar=1$ throughout).
$a_k$ is the annihilation operator of the $k$th oscillator, $\Ds[o]\rho=o\rho o^{\dagger}-\frac{1}{2}\{o^\dagger o,\rho\}$ is the dissipator, $\gamma_{\uparrow}$ and $\gamma_{\downarrow}$ are the pumping and nonlinear damping rates, $\kappa_{kj}=\kappa_{jk}\geq0$ is the dissipative coupling strength.
At the mean-field level, the competition between pumping and nonlinear loss stabilizes a limit cycle of amplitude $r_{\text{LC}}=\sqrt{\gamma_{\uparrow}/(2\gamma_{\downarrow})}$ in each oscillator, while the nonlocal dissipators add damping to the relative motion and thereby make the phases attract and lock~\cite{Walter2015,Walter2014a,Lee2014a}.

The nonlocal dissipative terms encode the coupling pattern among the vdP oscillators, which is most compactly described by a graph Laplacian~\cite{Pecora1998,Barahona2002,Arenas2008}.
For a connected undirected network, the couplings define the adjacency matrix $K_{kj}=\kappa_{kj}$ ($K_{kk}=0$) and the degree matrix $D=\mathrm{diag}(\sum_{l}\kappa_{kl})$, so that the nonlocal part of Eq.~\eqref{Eq:ME} is encoded entirely by the $N\times N$ matrix $L=D-K$ acting on the operator vector $\boldsymbol{a}=(a_{1},\ldots,a_{N})^{\mathrm T}$,
\begin{equation}
L=\sum_{\mu=1}^{N}\lambda_{\mu}\boldsymbol{u}_{\mu}\boldsymbol{u}_{\mu}^{\mathrm T},
\qquad
0=\lambda_{1}<\lambda_{2}\leq\cdots\leq\lambda_{N},
\label{Eq:Laplacian}
\end{equation}
with orthonormal eigenvectors $\boldsymbol{u}_{\mu}$.
The uniform eigenvector $\boldsymbol{u}_{1}\propto(1,\ldots,1)^{\mathrm T}$ is the collective in-phase mode and is immune to the dissipative coupling, while each transverse mode $\mu\geq2$ acquires an additional damping $\lambda_{\mu}/2$ in its mean-field amplitude (see Supplementary Material (SM)~\cite{supp}\nocite{Bergholm2022,Bond2024,Cattaneo2021a,Karpat2019,Li2026c,Walls2008}).
The pumping is diagonal in this basis as well, so that only the nonlinear loss couples the different modes.
Independently of the coupling, the collective background relaxes on the timescale $\gamma_{\uparrow}^{-1}$.
Which of these channels controls the synchronization error therefore depends on how the compared initial states populate them.
This structure holds for any $N$; we first address the minimal case of $N=2$, where the collective background, rather than a transverse mode, sets the late-time ordering.

\blueit{SMpE in two quantum vdP oscillators.--}
For two oscillators with $\kappa_{12}\equiv\kappa$, the graph Laplacian $L=\kappa\bigl(\begin{smallmatrix}1&-1\\-1&1\end{smallmatrix}\bigr)$ has the collective zero mode $\boldsymbol{u}_{1}=(1,1)^{\mathrm T}/\sqrt{2}$ and the transverse eigenvector $\boldsymbol{u}_{2}=(1,-1)^{\mathrm T}/\sqrt{2}$ with $\lambda_{2}=2\kappa$; the latter defines the relative mode $a_{-}=\boldsymbol{u}_{2}^{\mathrm T}\boldsymbol{a}=(a_{1}-a_{2})/\sqrt{2}$, damped by the coupling at the amplitude rate $\lambda_{2}/2=\kappa$~\cite{supp}.

To compare initial states with different degrees of synchronization, we adopt the measure introduced by~\citet{Mari2013},
\begin{equation}
\Ss_{c}(t)=\frac{1}{D_c(t)}\equiv\langle q_{-}(t)^{2}+p_{-}(t)^{2}\rangle^{-1},
\label{Eq:Sc}
\end{equation}
with the relative quadratures $q_{-}=(q_1-q_2)/\sqrt{2}$ and $p_{-}=(p_1-p_2)/\sqrt{2}$, where $q_k=(a_k+a_k^\dagger)/\sqrt{2}$ and $p_k=i(a_k^\dagger-a_k)/\sqrt{2}$.
The synchronization error $D_c(t)$ combines the mismatch of the mean quadratures with their quantum fluctuations; the uncertainty relation bounds it as $0<\Ss_c(t)\leq1$, with equality when the relative mode is in the vacuum.
\begin{figure}[!htpb]
    \centering
    \includegraphics[width=0.46\textwidth]{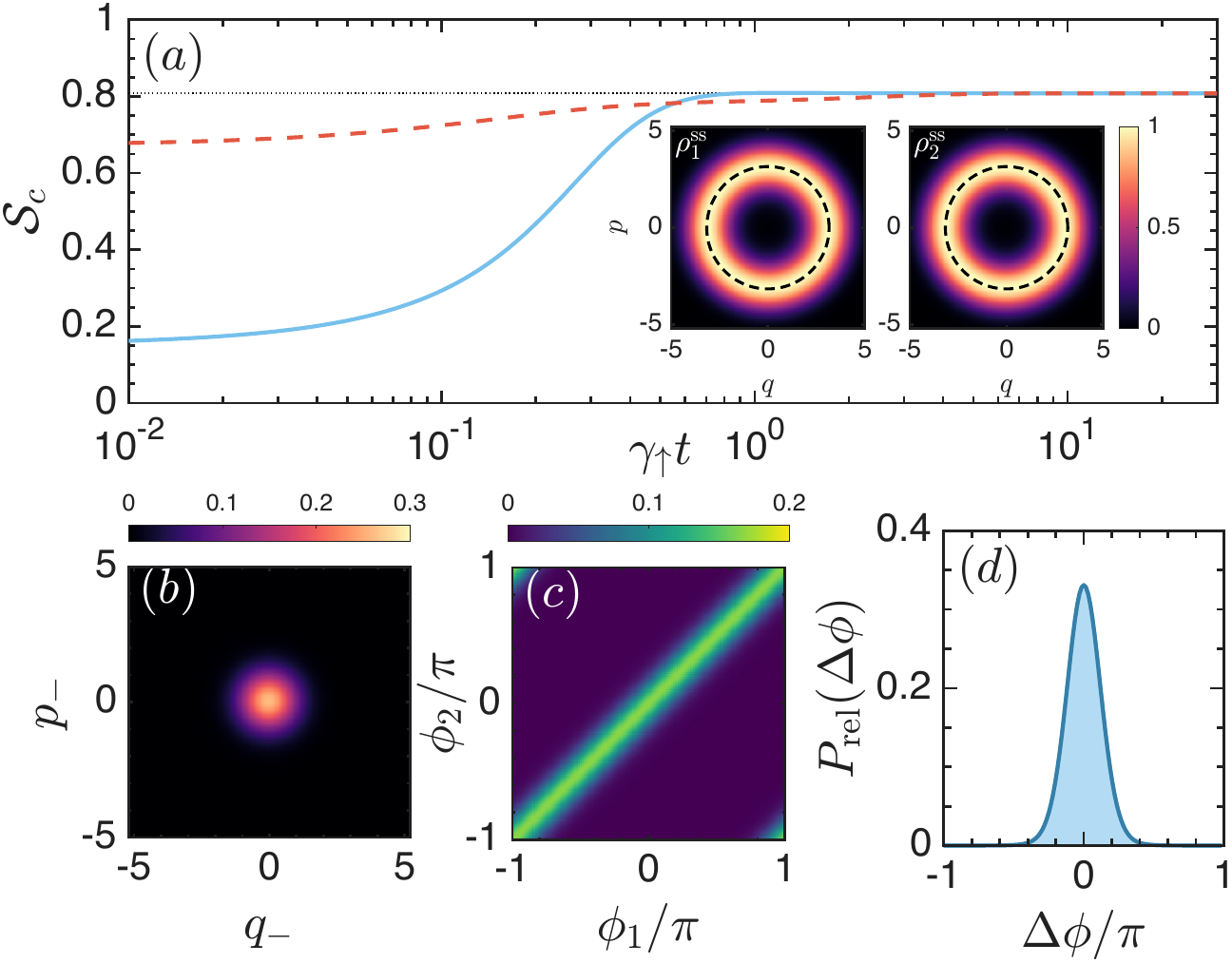}
    \caption{\label{Fig:N2Mpemba} SMpE in two dissipatively coupled quantum vdP oscillators.
    (a) Synchronization measure $\Ss_{c}(t)$ for $\rho_{A}$ (solid blue) and $\rho_{B}$ (dashed red), and  $\Ss_{c}^{A}(t)$ overtakes $\Ss_{c}^{B}(t)$ at $\gamma_\uparrow t_\times\simeq0.56$.
    Inset: normalized steady-state Wigner function $W(q,p)/\max{[W(q,p)]}$, exhibiting the ring structure of the limit cycle.
    The black dashed circles indicate the radius $r_{\text{LC}}$.
    (b)-(d) Characterization of the steady state $\rho_{\rm ss}$: (b)
    Relative-mode Wigner function $W(q_{-},p_{-})$ is concentrated at the origin, indicating a strongly suppressed synchronization error.
    (c) Joint phase distribution $P(\phi_{1},\phi_{2})$ forms a ridge along $\phi_{1}\simeq\phi_{2}$, indicating phase locking accompanied by global-phase diffusion.
    (d) Relative-phase distribution $P_{\rm rel}(\Delta\phi)$ is sharply peaked at $\Delta\phi=0$, confirming a phase-locked state.
    The parameters are $\omega=\gamma_{\uparrow}$, $\gamma_{\downarrow}=0.1\gamma_{\uparrow}$, and $\kappa=4\gamma_{\uparrow}$.}
\end{figure}

We initialize the two oscillators in product coherent states $\rho_X=|\alpha_{1,X},\alpha_{2,X}\rangle\langle\alpha_{1,X},\alpha_{2,X}|$, $X=A,B$, with $\alpha_{1,X}=\eta_{1,X}r_{\text{LC}}$ and $\alpha_{2,X}=\eta_{2,X}r_{\text{LC}}e^{i\Delta\phi_0}$. 
The initial phase difference is fixed to $\Delta\phi_0=\pi/4$ and $(\eta_{1},\eta_{2})_{A}=(1,\frac{3}{2})$, $(\eta_{1},\eta_{2})_{B}=(\frac{2}{5},\frac{3}{20})$.
For product coherent states, Eq.~\eqref{Eq:Sc} gives exactly $\Ss_c(0)=1/(1+|\alpha_1-\alpha_2|^2)$, so that $\Ss_c^{A}(0)=0.15$ and $\Ss_c^{B}(0)=0.67$.

\begin{figure}[!htpb]
    \centering
    \includegraphics[width=0.46\textwidth]{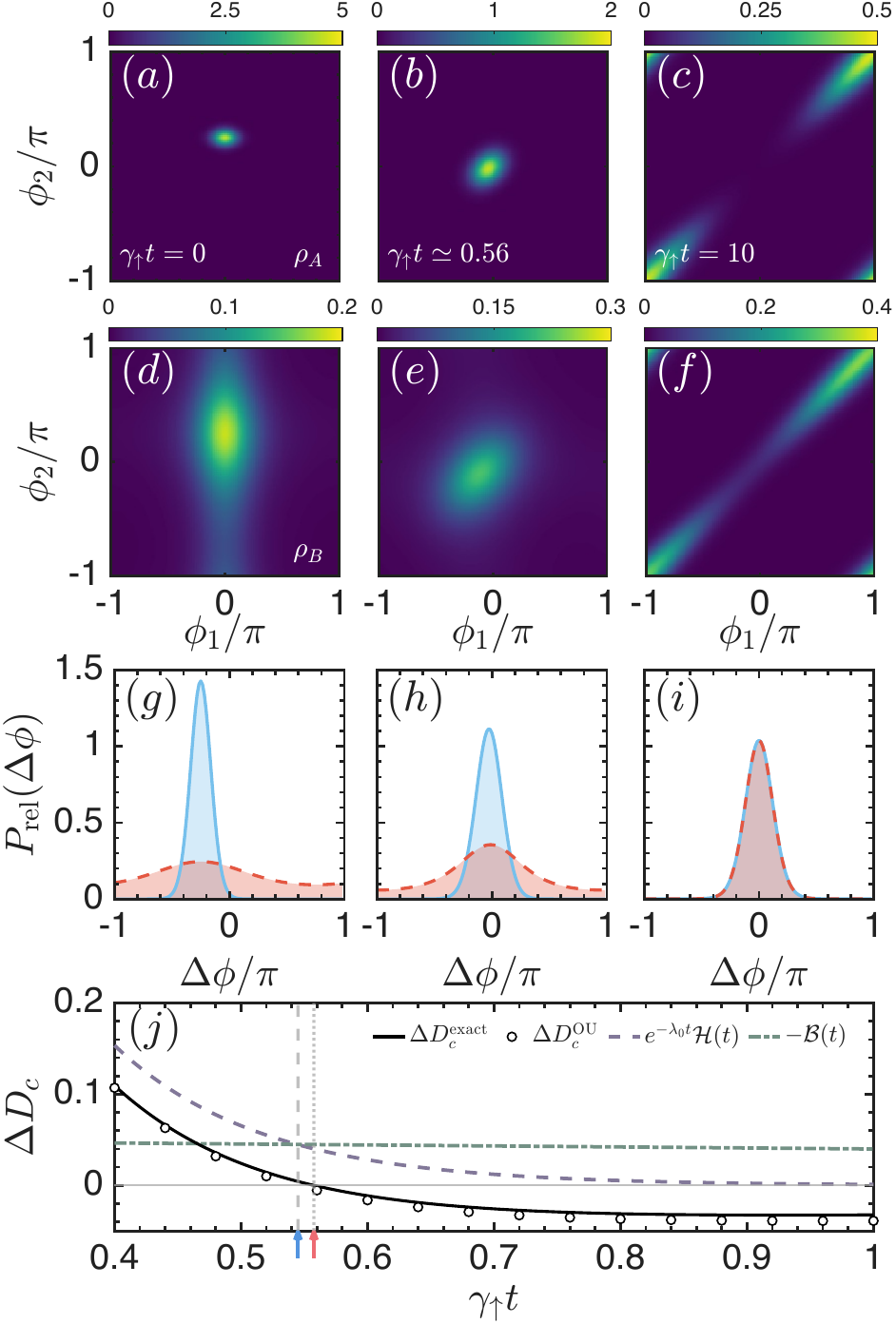}
    \caption{\label{Fig:N2Dyn} Phase dynamics and mechanism of the SMpE in two dissipatively coupled quantum vdP oscillators.
    (a)--(f) Joint phase distribution $P(\phi_1,\phi_2)$ for $\rho_A$ [(a)--(c)] and $\rho_B$ [(d)--(f)] at $\gamma_\uparrow t=0$ [(a),(d)], the crossing time $\gamma_\uparrow t_\times\simeq0.56$ [(b),(e)], and $\gamma_\uparrow t=10$ [(c),(f)].
    Both states evolve toward the diagonal ridge along $\phi_1\simeq\phi_2$ that signals phase locking.
    (g)--(i) Corresponding relative-phase distributions $P_{\mathrm{rel}}(\Delta\phi)$ for $\rho_A$ (blue solid) and $\rho_B$ (red dashed) at the same instants.
    The sharp initial peak of $\rho_A$ at the imprinted $\Delta\phi=-\pi/4$ is progressively dragged toward $\Delta\phi=0$, merging with the initially broad distribution of $\rho_B$ by $\gamma_\uparrow t=10$.
    (j) Difference of synchronization errors, $\Delta D_c(t)\equiv D_c^{A}(t)-D_c^{B}(t)$: exact result (solid line) versus OU description (circles), together with the slowly decaying background $-\Bs(t)$ (dash-dotted) and memory term $e^{-\lambda_{0}t}\Hs(t)$ (dashed); arrows mark $t_{\times}$ from Eq.~\eqref{Eq:N2-crossing} (blue, $\gamma_\uparrow t_\times\simeq0.55$) and exact master equation (red, $\gamma_\uparrow t_\times\simeq0.56$).
    The parameters are the same as in Fig.~\ref{Fig:N2Mpemba}.}
\end{figure}

Nevertheless, $\rho_A$ synchronizes faster.
Figure~\ref{Fig:N2Mpemba}(a) shows $\Ss_c(t)$ for $\rho_A$ (solid blue) and $\rho_B$ (dashed red)~\cite{N2note}.
The initially lower curve $\Ss_c^{A}(t)$ grows rapidly and overtakes $\Ss_c^{B}(t)$ at the crossing time $t_\times$ ($\gamma_\uparrow t_\times\simeq0.56$), before both curves converge toward the steady state; this overtaking of the initially better synchronized state constitutes the SMpE.
However, a large value of $\Ss_c$ alone is not sufficient evidence of synchronization: the vacuum state even yields $\Ss_c=1$ without any limit cycle or phase locking.
To certify synchronization we therefore examine $\rho_{\rm ss}$ itself.
The limit cycle is manifest in single-oscillator Wigner functions, which form rings at $|\alpha|=r_{\text{LC}}$ [Fig.~\ref{Fig:N2Mpemba}(a), inset], reflecting a well-defined amplitude but an uncertain phase.
The relative-mode Wigner function $W(q_{-},p_{-})$ maps synchronization error into phase space; its concentration at the origin [Fig.~\ref{Fig:N2Mpemba}(b)] shows that the error is strongly suppressed.
Phase locking is revealed by the phase statistics of $\rho_{\rm ss}$, obtained from the canonical phase POVM in the Fock space~\cite{Barak2005,Hush2015,Davis-Tilley2016},
\begin{equation}
P(\phi_1,\phi_2)=\frac{1}{(2\pi)^2}
\sum_{n_1,n_2,m_1,m_2}
\rho_{n_1n_2,m_1m_2}
e^{i(m_1-n_1)\phi_1}
e^{i(m_2-n_2)\phi_2},
\end{equation}
where $\rho_{n_1n_2,m_1m_2}=\langle n_{1},n_{2}|\rho_{\text{ss}}|m_{1},m_{2}\rangle$ denotes $\rho_{\text{ss}}$ in the Fock basis $|n_1,n_2\rangle=|n_1\rangle\otimes|n_2\rangle$.
The joint phase distribution forms a ridge along $\phi_1\simeq\phi_2$ [Fig.~\ref{Fig:N2Mpemba}(c)]: the two phases are locked, while the collective phase diffuses freely, in line with the $U(1)$ symmetry of Eq.~\eqref{Eq:ME}.
Accordingly, the relative-phase distribution $P_{\rm rel}(\Delta\phi)=\int_{-\pi}^{\pi}P(\phi+\Delta\phi,\phi)\,d\phi$, with $\Delta\phi=\phi_1-\phi_2$, is sharply peaked at $\Delta\phi=0$~\cite{Jessop2020} [Fig.~\ref{Fig:N2Mpemba}(d)], demonstrating that the two oscillators settle into a phase-locked state with near-in-phase synchronization.

Both states relax to the same synchronized steady state, so the crossing must originate in their transients, which we now resolve in phase space.
In Figs.~\ref{Fig:N2Dyn}(a)--(f), joint phase distributions evolve from opposite characters toward the same phase-locked ridge, with $\rho_{A}$ starting as a localized bump of well-defined phase and $\rho_{B}$ broadly spread deep inside the limit cycle, its phase poorly defined.
Strikingly, in relative-phase distributions [Figs.~\ref{Fig:N2Dyn}(g)--(i)], $\rho_{A}$'s peak at $t_{\times}$ is already markedly sharper than $\rho_{B}$'s [Fig.~\ref{Fig:N2Dyn}(h)].
Were phase locking the whole content of $\Ss_{c}$, this sharper peak would imply the larger $\Ss_{c}^{A}$, contradicting the crossing.
The joint distribution [Fig.~\ref{Fig:N2Dyn}(b)] resolves the paradox, showing that $\rho_{A}$ remains displaced from the diagonal, so its error is carried by relative-mode displacement, whereas that of $\rho_{B}$ is carried by phase fluctuations.
The crossing is thus an exchange of dominance between these two error components, not a convergence of the states, which remain visibly distinct at $t_{\times}$.

This exchange can be quantified, since $\Ss_c$ depends only on the first and second moments of $a_{-}$ [cf. Eq.~\eqref{Eq:Sc}], whose equations close if the mode remains Gaussian.
The nonlinear loss that sustains the limit cycle makes the full state non-Gaussian, yet a reduced mode can remain near Gaussian, since pumping and nonlocal coupling are linear jump operators and only the loss can inject non-Gaussianity into $\rho_{-}$.
The non-Gaussianity of $\rho_{-}$ remains below $1.5\times10^{-4}$ for both states (see End Matter), justifying closed second-moment equations and an effective Ornstein--Uhlenbeck (OU) description (see Ref.~\cite{supp}).

Within this Gaussian closure we linearize the relative-mode dynamics about the instantaneous collective state: since the coupling is diagonal in the graph Laplacian basis and the noise is uniform on the synchronized background, each transverse mode $\mu$ obeys an independent Langevin equation for its amplitude mismatch $x_{\mu}$ and phase error $y_{\mu}$, with drift and diffusion coefficients set by the collective occupation $n_{1}(t)$~\cite{supp}.
Here $n_{1}(t)$ obeys the logistic equation $\dot n_{1}=\gamma_{\uparrow}n_{1}-(2\gamma_{\downarrow}/N)n_{1}^{2}$ and relaxes to its stationary value $n_{1,\text{ss}}$, with $z(t)=n_{1}(t)/n_{1,\text{ss}}$ starting from $z(0)=1/(1+c)$ and approaching unity, where $c$ is fixed by the initial occupation.
The variances $X_{\mu}=\langle x_{\mu}^{2}\rangle$ and $Y_{\mu}=\langle y_{\mu}^{2}\rangle$ obey $\dot X_{\mu}=-[\lambda_{0,\mu}+3\gamma_{\uparrow}z(t)]X_{\mu} +q_{0,\mu}+\gamma_{\uparrow}z(t)$ and $\dot Y_{\mu}=-[\lambda_{0,\mu}+\gamma_{\uparrow}z(t)]Y_{\mu} +q_{0,\mu}+\gamma_{\uparrow}z(t)$, with $q_{0,\mu}=(\lambda_{\mu}+\gamma_{\uparrow})/2$ and $\lambda_{0,\mu}=\lambda_{\mu}-\gamma_{\uparrow}$.
The variance equations are linear with coefficients set by $z(t)$, since each solution splits exactly into a particular part, a polynomial in $z(t)$, and a homogeneous part $e^{-\lambda_{0,\mu}t}$ times a slowly varying memory factor.
For each transverse mode, the difference between two states therefore separates into a background difference $\Bs_\mu(t)=\sum_{n=1,3}\bigl(P_{n,\mu}[z_A(t)]-P_{n,\mu}[z_B(t)]\bigr)$ and a memory term $\Hs_\mu(t)=\sum_{n=1,3}\bigl(\bigl[M^A_{\mu,n}(0)-P_{n,\mu}[z_A(0)]\bigr]R_A^{n}(t) -\bigl[M^B_{\mu,n}(0)-P_{n,\mu}[z_B(0)]\bigr]R_B^{n}(t)\bigr)$, where $P_{n,\mu}[z]$ is the polynomial particular solution of the variance equations, $R(t)$ is the memory factor with $R(0)=1$, $M^{A,B}_{\mu,n}(0)$ are the initial variances, with $M_{\mu,3}=X_\mu$ and $M_{\mu,1}=Y_\mu$.
For $N=2$ the sole transverse mode is $\mu=2$, we have $D_c(t)=X(t)+Y(t)$, so the difference between the two synchronization errors is~\cite{supp}
\begin{equation}
\Delta D_c(t)=\Bs(t)+e^{-\lambda_{0}t}\Hs(t),
\qquad
\lambda_{0}\equiv\lambda_{0,2}=2\kappa-\gamma_{\uparrow},
\label{Eq:DeltaDc}
\end{equation}
with $\Bs\equiv\Bs_2$ and $\Hs\equiv\Hs_2$.
The SMpE occurs at $\Delta D_c(t_\times)=0$.
Writing $\ell(t)\equiv\ln|\Hs(t)/\Bs(t)|=\ell_0+\ell_1\gamma_{\uparrow}t +\tfrac{1}{2}\ell_2(\gamma_{\uparrow}t)^{2}+\cdots$; the crossing condition $\ell(t_\times)=\lambda_0t_\times$ gives
\begin{equation}
t_\times\simeq
\frac{(\lambda_0/\gamma_{\uparrow}-\ell_1)
-\sqrt{(\lambda_0/\gamma_{\uparrow}-\ell_1)^{2}-2\ell_2\ell_0}}{\ell_2},
\label{Eq:N2-crossing}
\end{equation}
where $\ell_{0,1,2}$ are the coefficients of the expansion~\cite{supp}.

\begin{figure}[!htpb]
    \centering
    \includegraphics[width=0.46\textwidth]{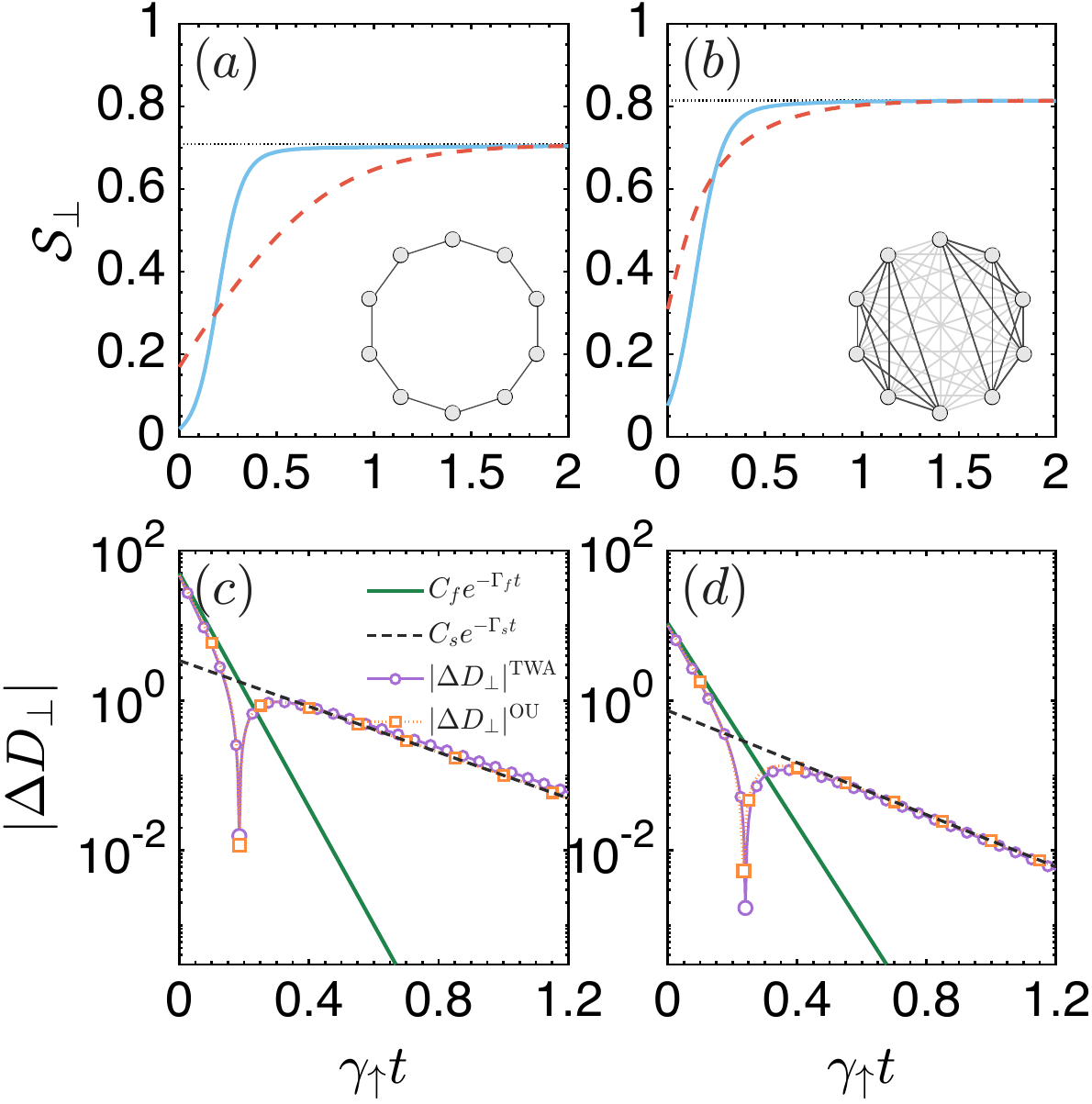}
    \caption{\label{Fig:N10Mpemba} SMpE in networks of $N=10$ dissipatively coupled quantum vdP oscillators.
    (a), (b) Synchronization measure $\Ss_{\perp}(t)$ for $\rho_{A}$ (solid blue) and $\rho_{B}$ (dashed red) in (a) a uniform ring and (b) a two-cluster all-to-all network with couplings $\kappa_{\rm in}$ (intra-cluster) and $\kappa_{\rm out}$ (inter-cluster); insets show the coupling graphs.
    (c), (d) Mode competition in (c) the ring and (d) the two-cluster network: the fast (solid green) and slow (dashed black) contributions $C_{f}e^{-\Gamma_{f}t}$ and $C_{s}e^{-\Gamma_{s}t}$, with $\Gamma_{f,s}=\lambda_{f,s}+2\gamma_{\uparrow}$, to $\Delta D_{\perp}(t)$, compared with the TWA result $|\Delta D_{\perp}|^{\mathrm{TWA}}$ (circles) and OU description $|\Delta D_{\perp}|^{\mathrm{OU}}$ (squares).
    The crossings occur at $\gamma_\uparrow t_\times\approx0.184$ (ring) and $0.241$ (two-cluster) in the TWA, compared with the OU description $0.185$ and $0.233$.
    The parameters are $\omega=\gamma_{\uparrow}$, $\gamma_{\downarrow}=0.05\gamma_{\uparrow}$, and $C_f/C_s\approx14.6$; the ring has $\kappa=4\gamma_{\uparrow}$ and the two-cluster network has $\kappa_{\rm in}=2.5\gamma_{\uparrow}$ and $\kappa_{\rm out}=0.2\gamma_{\uparrow}$.}
\end{figure}

Figure~\ref{Fig:N2Dyn}(j) compares the exact and OU results for $\Delta D_c(t)$: the OU circles follow the exact curve (solid line) throughout, and the two crossing times almost coincide.
The same panel resolves $\Delta D_c(t)$ into its two terms: the memory term $e^{-\lambda_0t}\Hs(t)$ (dashed) decays rapidly and governs the early dynamics, whereas the slowly varying background (dash-dotted) persists and fixes the late-time approach.
The crossing occurs where the two balance.
Moreover, Eq.~\eqref{Eq:DeltaDc} gives a sufficient condition for the SMpE.
For $\lambda_0>0$ the memory term is asymptotically subleading, so $\Delta D_c(t)\simeq Q'[1](c_B-c_A)e^{-\gamma_{\uparrow}t}$ with $Q[z]\equiv P_{3}[z]+P_{1}[z]$; since $\Delta D_c(0)>0$, a crossing is guaranteed by continuity provided $Q'[1](c_B-c_A)<0$~\cite{supp}.


\blueit{SMpE in quantum vdP oscillator networks.--}
For a network, we extend the measure of Eq.~\eqref{Eq:Sc} to $N$ oscillators as~\cite{supp}
\begin{equation}
\Ss_{\perp}(t)=\frac{1}{D_{\perp}(t)}\equiv\biggl[\frac{1}{N-1}\sum_{\mu=2}^{N}
\bigl\langle Q_{\mu}(t)^{2}+P_{\mu}(t)^{2}\bigr\rangle\biggr]^{-1},
\label{Eq:Sperp}
\end{equation}
where $Q_{\mu}=\boldsymbol{u}_{\mu}^{\mathrm T}\boldsymbol{q}$ and $P_{\mu}=\boldsymbol{u}_{\mu}^{\mathrm T}\boldsymbol{p}$ are the quadratures of the transverse mode $\mu$, with $\{\boldsymbol{u}_{\mu}\}$ the eigenbasis of $L$ [Eq.~\eqref{Eq:Laplacian}], $\boldsymbol{q}=(q_{1},\ldots,q_{N})^{\mathrm T}$, $\boldsymbol{p}=(p_{1},\ldots,p_{N})^{\mathrm T}$, and $q_{k},p_{k}$ the single-oscillator quadratures.
The measure is bounded as $0<\Ss_{\perp}\le1$, for $N=2$ it reduces to the measure of Eq.~\eqref{Eq:Sc}.

\begin{figure}[!htpb]
    \centering
    \includegraphics[width=0.46\textwidth]{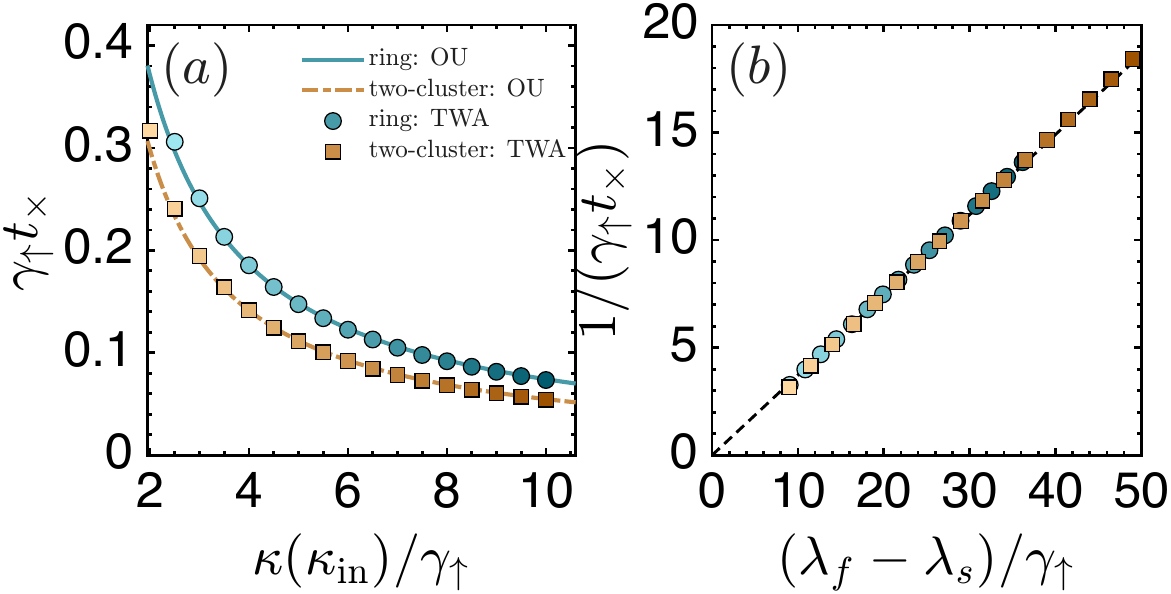}
    \caption{\label{Fig:N10laplacian} SMpE crossing time dictated by the graph-Laplacian spectral gap. 
    (a) The $\gamma_\uparrow t_\times$ versus the coupling for a uniform ring ($\kappa$) and a two-cluster network ($\kappa_{\rm in}$, with $\kappa_{\rm out}=0.2\gamma_\uparrow$ fixed).
    The lines are the OU description, with $C_f/C_s\approx14.6$ fixed for both networks; circles (ring) and squares (two-cluster) are the TWA results.
    (b) The inverse crossing time of both networks versus the spectral gap $(\lambda_f-\lambda_s)/\gamma_\uparrow$ collapse onto a line (dashed).
    The parameters are $\omega=\gamma_\uparrow$ and $\gamma_\downarrow=0.05\gamma_\uparrow$.}
\end{figure}

In the mean-field description, the coupling damps each transverse mode at $\lambda_{\mu}/2$ while the gain drives it at $\gamma_{\uparrow}/2$, so the synchronized manifold is attractive whenever $\lambda_{2}>\gamma_{\uparrow}$.
Unlike the $N=2$ case, where a single transverse channel is available, a network provides $N-1$ transverse channels resolved by the eigenvectors of $L$, with the slowest populated one setting the relaxation bottleneck.
The SMpE can therefore be designed directly from the Laplacian eigenbasis: choosing which modes carry the excess initial error selects the channels that relax it, and hence the crossing.

To test this design principle, we consider a network of dissipatively coupled vdP oscillators.
Since the many-body Hilbert space grows exponentially with $N$, we evaluate Eq.~\eqref{Eq:ME} within the truncated-Wigner approximation (TWA)~\cite{Polkovnikov2010,supp,TWAnote}.
We take $N=10$ oscillators and consider two topologies: a uniform ring with $\kappa=4\gamma_{\uparrow}$, and a network of two all-to-all clusters, with intra- and inter-cluster couplings $\kappa_{\rm in}=2.5\gamma_{\uparrow}$ and $\kappa_{\rm out}=0.2\gamma_{\uparrow}$ [insets of Fig.~\ref{Fig:N10Mpemba}(a),(b)].
Both Laplacian spectra are known analytically: for the ring, $\lambda_{k}=2\kappa[1-\cos(2\pi k/N)]$, so that the slowest and fastest transverse modes are $\lambda_{s}=2\kappa[1-\cos(\pi/5)]$ and $\lambda_{f}=4\kappa$; for the two-cluster network, $\lambda_{s}=10\kappa_{\rm out}$ and $\lambda_{f}=5(\kappa_{\rm in}+\kappa_{\rm out})$.

For each topology we prepare $\rho_{A}$ and $\rho_{B}$, as $\alpha_{k}(0)=r_{\mathrm{LC}}+\xi_{k}+\delta_{k}$, with thermal Wigner noise $\xi_{k}$ of occupation $n_{\mathrm{th}}=0.75$ and deterministic transverse displacements $\boldsymbol{\delta}_{A}=a_{f}\boldsymbol{u}_{f}$ and $\boldsymbol{\delta}_{B}=a_{s}\boldsymbol{u}_{s}$ along normalized fastest and slowest transverse eigenvectors, respectively.
These displacements raise the excess initial errors by $C_{f}=2a_{f}^{2}/(N-1)$ and $C_{s}=2a_{s}^{2}/(N-1)$, respectively, so that $\Delta D_{\perp}(0)\equiv D_{\perp}^{A}(0)-D_{\perp}^{B}(0)=C_{f}-C_{s}$; fixing $C_{f}/C_{s}\approx14.6$ for every network makes $\rho_{A}$ the initially less synchronized state.

Figures~\ref{Fig:N10Mpemba}(a)-(b) show the resulting $\Ss_{\perp}(t)$ for the two topologies.
In both cases the initially less synchronized state $\rho_{A}$ rises rapidly and overtakes $\rho_{B}$ at $\gamma_{\uparrow}t_{\times}=0.184$ (ring) and $0.241$ (two-cluster network), exhibiting the SMpE.
Because $\rho_{A}$ and $\rho_{B}$ share the same collective background ($c_{A}=c_{B}=0$), the background terms cancel exactly, and the difference between the two errors is carried by the transverse modes alone, $\Delta D_{\perp}(t)=C_{f}e^{-\Gamma_{f}t}-C_{s}e^{-\Gamma_{s}t}$ with $\Gamma_{\mu}=\lambda_{\mu}+2\gamma_{\uparrow}$.
The chosen initial states guarantee a crossing: $C_{f}>C_{s}$ makes $\Delta D_{\perp}(0)>0$, while $\Gamma_{f}>\Gamma_{s}$ makes $\Delta D_{\perp}(t)$ negative at late times, so the inversion of $\Ss_{\perp}$ is built in.
Panels (c),(d) show $|\Delta D_{\perp}(t)|$ for the TWA (circles) and OU description (squares).
The early dynamics follows the fast contribution $C_{f}e^{-\Gamma_{f}t}$ and the subsequent evolution the slow one $C_{s}e^{-\Gamma_{s}t}$, the two crossing at $t_{\times}$.
Setting $\Delta D_{\perp}(t_{\times})=0$ gives
\begin{equation}
t_{\times}=\frac{\ln(C_{f}/C_{s})}{\lambda_{f}-\lambda_{s}},
\label{Eq:Unified-crossing}
\end{equation}
which yields $\gamma_{\uparrow}t_{\times}=0.185$ and $0.233$ for the two networks, in good agreement with the TWA results.

Beyond this single parameter set, Fig.~\ref{Fig:N10laplacian} shows that, as the coupling is varied, the crossing time is independent of the network topology (see End Matter).
Panel (a) compares both topologies with the curves of Eq.~\eqref{Eq:Unified-crossing} evaluated from the spectra above ($t_{\times}\propto1/\kappa$; $t_{\times}\propto1/(\kappa_{\rm in}-\kappa_{\rm out})$), which the TWA results follow across the whole scan.
Panel (b) shows that the two networks obey one and the same relation: plotted against the spectral gap, the crossing times fall onto a single line through the origin, whose common slope $1/\ln(C_f/C_s)$ is fixed by the preparation.
The graph-Laplacian spectrum thus decides, from the initial states and the system parameters alone, both whether the SMpE occurs and when~\cite{supp}, so that the acceleration can be designed and predicted without solving the many-body dynamics.

\blueit{Conclusions.--} We demonstrated the SMpE, an anomalous relaxation in which an initially less synchronized state overtakes a more synchronized one as both relax to the same phase-locked attractor.
We developed a graph-Laplacian analysis of the SMpE and extended it to quantum many-body models with arbitrary nonlocal dissipative couplings~\cite{Cabot2019,Cabot2021,Wachtler2023,Wachtler2024a}, where the synchronization dynamics can be designed and controlled through the Laplacian matrix rather than the exponentially large Liouvillian. 
Experimentally, quantum vdP oscillator synchronization has been realized with trapped ions~\cite{Liu2026n}; we also propose a quantum-circuit implementation based on the collision model~\cite{Ciccarello2022,Li2024b,Cattaneo2021,supp}.
Our work paves the way for controlling synchronization dynamics in quantum many-body systems and for exploiting quantum synchronization as a resource.

\blueit{Acknowledgments.--} This work was supported by the National Natural Science Foundation of China under Grant (No. 12605045, No. 12605041, No. 12304389), the fundamental research funds for the central universities (Grant No. GK202601011), and the Scientific Research Foundation of NEU (Grant No. 01270021920501*115).
X. L. also acknowledges support from the National Natural Science Foundation of China (Grant No. 12247101), the Fundamental Research Funds for the Central Universities (Grant No.lzujbky-2025-jdzx07), the Natural Science Foundation of Gansu Province (No. 22JR5RA389, No.25JRRA799) and the ‘111 Center’ under Grant No. B20063.

\blueit{Data availability.--}
Data supporting this article are openly available~\cite{li_2026_23879363}.

\resetappendix

\section{End Matter}

\begin{figure}[!htpb]
    \centering
    \includegraphics[width=0.46\textwidth]{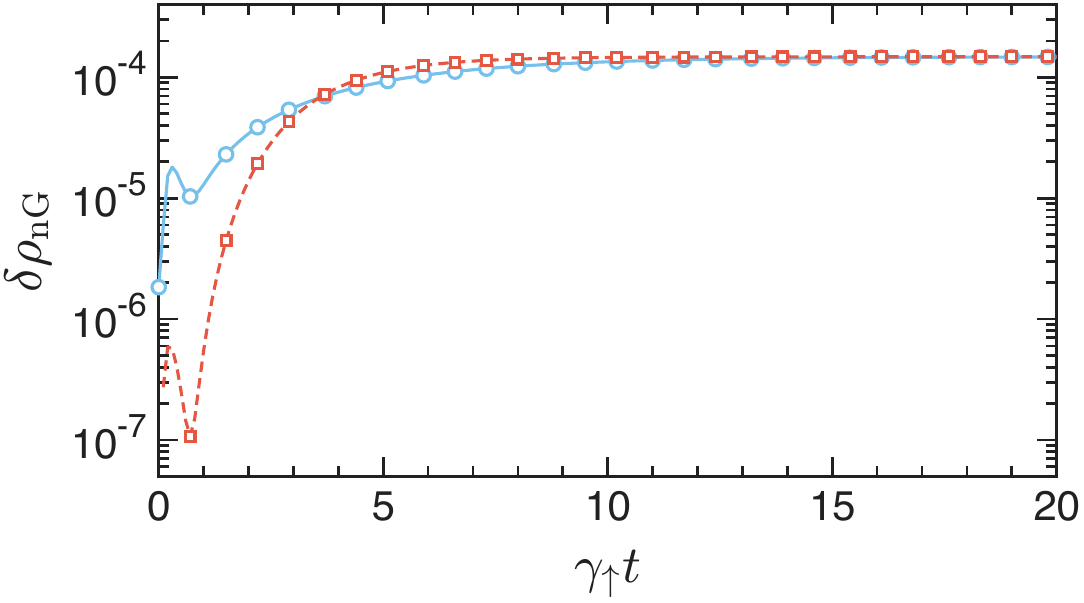}
    \caption{\label{Fig:N2NG} Non-Gaussianity $\delta\rho_{\mathrm{nG}}(t)$ of the relative mode for the two initial states used in Fig.~\ref{Fig:N2Mpemba}, $\rho_{A}$ (blue circles, solid line) and $\rho_{B}$ (red squares, dashed line).
    The parameters are the same as in Fig.~\ref{Fig:N2Mpemba}.}
\end{figure}

\blueit{Appendix: Gaussianity of the relative mode.--} 
For the two-oscillator case, the relative mode is obtained by tracing out the collective mode, $\rho_{-}=\mathrm{Tr}_{+}[\rho]$.
Its non-Gaussianity is quantified by the relative entropy to the Gaussian reference state $\tau_{-}$ that shares the same first and second moments as
$\rho_{-}$~\cite{Genoni2008},
\begin{equation}
\delta\rho_{\mathrm{nG}}(t)\equiv S(\rho_{-}\Vert\tau_{-})=S(\tau_{-})-S(\rho_{-}),
\label{Eq:nG}
\end{equation}
with $S(\rho)=-\mathrm{Tr}[\rho\ln\rho]$ the von Neumann entropy, and $\delta\rho_{\mathrm{nG}}=0$ if and only if $\rho_{-}$ is Gaussian.
For both initial states of Fig.~\ref{Fig:N2Mpemba}, $\delta\rho_{\mathrm{nG}}(t)$ stays at or below $1.5\times10^{-4}$ throughout the evolution, including at the crossing (Fig.~\ref{Fig:N2NG}). 
This is consistent with the dynamics: the initial product coherent states leave the relative mode in a coherent state, and only the nonlinear loss can generate non-Gaussianity, whereas the nonlocal coupling $\kappa\Ds[a_{1}-a_{2}]$ is linear and Gaussianity preserving. 
The relative mode thus remains Gaussian to this accuracy, and the closed OU description built on it reproduces the exact dynamics.

\begin{figure}[!htpb]
    \centering
    \includegraphics[width=0.46\textwidth]{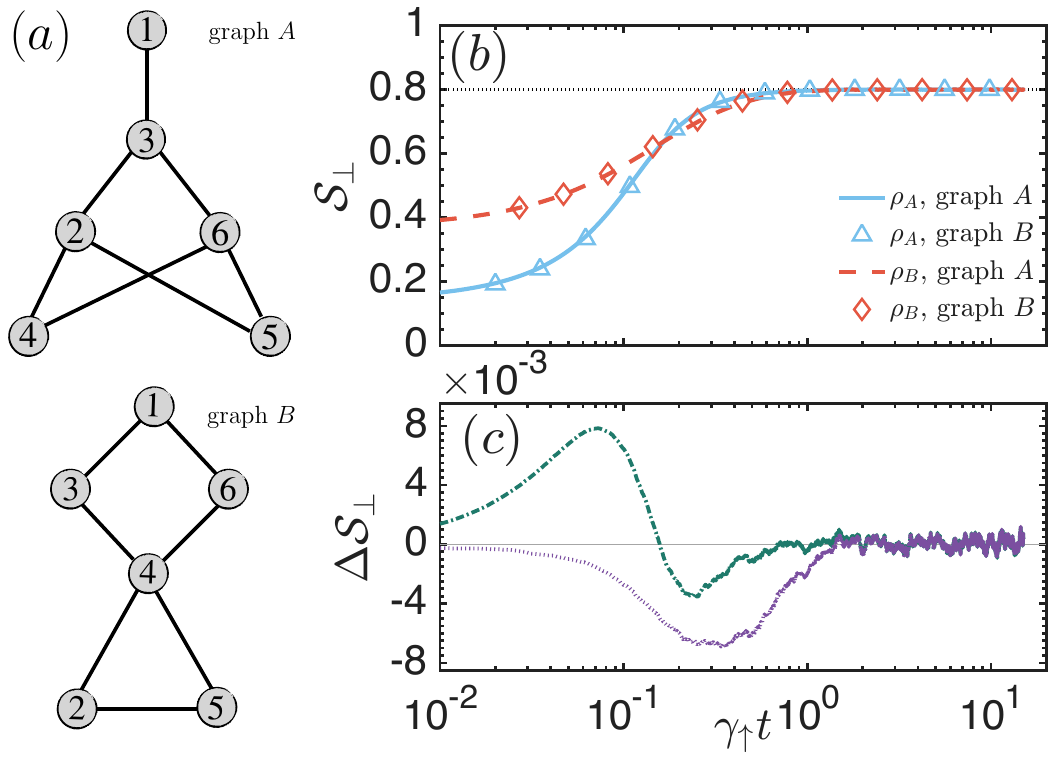}
    \caption{\label{Fig:N6} SMpE in two non-isomorphic but Laplacian-cospectral $N=6$ networks.
   (a) Graphs $A$ (top) and $B$ (bottom), with the common Laplacian spectrum $\kappa\{0,3-\sqrt5,2,3,3,3+\sqrt5\}$.
   (b) Synchronization measure $\Ss_{\perp}(t)$ for $\rho_{A}$ (blue) and $\rho_{B}$ (red) on graph $A$ (lines) and graph $B$ (markers); the crossings are $\gamma_{\uparrow}t_{\times}=0.165$ and $0.171$ for graphs $A$ and $B$, compared with the OU prediction $0.172$.
   (c) Difference $\Delta\Ss_{\perp}=\Ss_{\perp}(\text{graph }A)-\Ss_{\perp}(\text{graph }B)$ for $\rho_{A}$ and $\rho_{B}$.
   The parameters are $\omega=\gamma_{\uparrow}$, $\gamma_{\downarrow}=0.05\gamma_{\uparrow}$, $\kappa=4\gamma_{\uparrow}$, and $C_{f}/C_{s}\approx 21.8$.}
\end{figure}

\blueit{Appendix: Topology independence of the SMpE.--}
To show that the SMpE is not tied to a particular network structure, we compare two non-isomorphic $N=6$ graphs~\cite{Brouwer2012}.
Their network links are shown in Fig.~\ref{Fig:N6}(a), and their graph Laplacians are
\begin{equation}
  L_A = \kappa
  \begin{pmatrix}
     1 &  0 & -1 &  0 &  0 &  0\\
     0 &  3 & -1 & -1 & -1 &  0\\
    -1 & -1 &  3 &  0 &  0 & -1\\
     0 & -1 &  0 &  2 &  0 & -1\\
     0 & -1 &  0 &  0 &  2 & -1\\
     0 &  0 & -1 & -1 & -1 &  3
  \end{pmatrix},
\end{equation}
and
\begin{equation}
  L_B =\kappa
  \begin{pmatrix}
     2 &  0 & -1 &  0 &  0 & -1\\
     0 &  2 &  0 & -1 & -1 &  0\\
    -1 &  0 &  2 & -1 &  0 &  0\\
     0 & -1 & -1 &  4 & -1 & -1\\
     0 & -1 &  0 & -1 &  2 &  0\\
    -1 &  0 &  0 & -1 &  0 &  2
  \end{pmatrix}.
\end{equation}
They share the spectrum $\kappa\{0,3-\sqrt5,2,3,3,3+\sqrt5\}$.
Here $\lambda_{s}=\kappa(3-\sqrt5)$ and $\lambda_{f}=\kappa(3+\sqrt5)$ are the slowest and fastest transverse Laplacian eigenvalues.
The initial states are prepared as for the $N=10$ networks, with $C_{f}/C_{s}\approx21.8$.
Figure~\ref{Fig:N6}(b) shows the TWA results of $\Ss_{\perp}(t)$ for $\rho_{A}$ and $\rho_{B}$ on the two graphs: the results nearly coincide and both exhibit the SMpE.
Moreover, we show the difference between the two graphs for each state in Fig.~\ref{Fig:N6}(c); it remains at the $10^{-3}$ level, so the two graphs produce the same synchronization dynamics.
Since the graphs differ only in their networks while sharing the same Laplacian spectrum, the SMpE is governed by the spectrum of $L$ rather than by the specific topology; our Laplacian framework thus captures a general property of dissipatively coupled oscillator networks.

\bibliography{filterRef}

\end{document}


\beginsupplement

\title{Supplementary Material for ``Synchronization Mpemba Effect"}

\author{Yan Li\orcidlink{0009-0007-7218-133X}}
\affiliation{School of Physics and Information Technology, Shaanxi Normal University, Xi’an, 710062, P. R. China}

\author{Wenlin Li\orcidlink{0000-0003-2032-0530}}
\affiliation{College of Sciences, Northeastern University, Shenyang 110819, China}
\affiliation{Research Center for Quantum Physics, Northeastern University, Shenyang 110819, China}

\author{Xingli Li\orcidlink{0000-0003-2339-6557}}
\email{lixingli0130@gmail.com}
\affiliation{Department of Physics, The Chinese University of Hong Kong, Shatin, New Territories, Hong Kong, China}
\affiliation{The Chinese University of Hong Kong Shenzhen Research Institute, Shenzhen 518057, China}
\affiliation{Lanzhou Center for Theoretical Physics, Key Laboratory of Theoretical Physics of Gansu Province, Key Laboratory of Quantum Theory and Applications of MoE, Gansu Provincial Research Center for Basic Disciplines of Quantum Physics, Lanzhou University, Lanzhou 730000, China}

\begin{abstract}

\end{abstract}
\date{\today}

\maketitle

\tableofcontents

\section{Quantum van der Pol oscillator network and the graph Laplacian}
In this section, we provide a detailed derivation of the master equation for the quantum van der Pol (vdP) oscillator network and its connection to the graph Laplacian.
The system consists of $N$ dissipatively coupled quantum vdP oscillators.
The Hamiltonian is given by $H=\omega\sum_{k}a_k^\dagger a_k$, where $\omega$ is the frequency of oscillator.
The dynamics is governed by the Lindblad master equation
\begin{equation}
\frac{d}{dt}\rho = -i[H, \rho] + \sum_{k=1}^{N}\left(\gamma_{\uparrow}\Ds[a^{\dagger}_{k}]\rho+\gamma_{\downarrow}\Ds[a_{k}^{2}]\rho\right)+\sum_{k<j}\kappa_{kj}\Ds[a_{k}-a_{j}]\rho,
\label{Eq:ME}
\end{equation}
where $\Ds[o]\rho = o\rho o^{\dagger}-\frac{1}{2}\{o^{\dagger}o, \rho\}$ is the dissipator.
The positive rates $\gamma_{\uparrow}$, $\gamma_{\downarrow}$ and $\kappa_{kj}$ represent the pumping, the nonlinear loss and the dissipative coupling strength between the $k$th and $j$th oscillators, respectively.
For simplicity, we assume that all oscillators have the same frequency $\omega_{k}=\omega$ for all $k$.

The last term in Eq.~\eqref{Eq:ME} describes nonlocal dissipation, which provides a dissipative mechanism for synchronizing the oscillators.
The jump operator $a_k-a_j$ acts collectively on two vertices of the network and selectively damps their relative (antisymmetric) mode, while leaving the in-phase mode dark with respect to this channel.
This suppresses differences between the oscillator amplitudes and thus promotes synchronization.
The topology of these pairwise dissipative couplings can be compactly encoded by a graph Laplacian.
For a network of $N$ oscillators, we introduce the weighted adjacency matrix $K\in\mathbb{R}^{N\times N}$, which describes the topology and strength of the pairwise dissipative couplings.
Its elements are defined as
\begin{equation}
K_{kj}=
\begin{cases}
\kappa_{kj}, & k \text{ and } j \text{ are coupled},\\
0, & \text{otherwise}.
\end{cases}
\end{equation}
\begin{figure}[!htpb]
    \centering
    \includegraphics[width=0.95\textwidth]{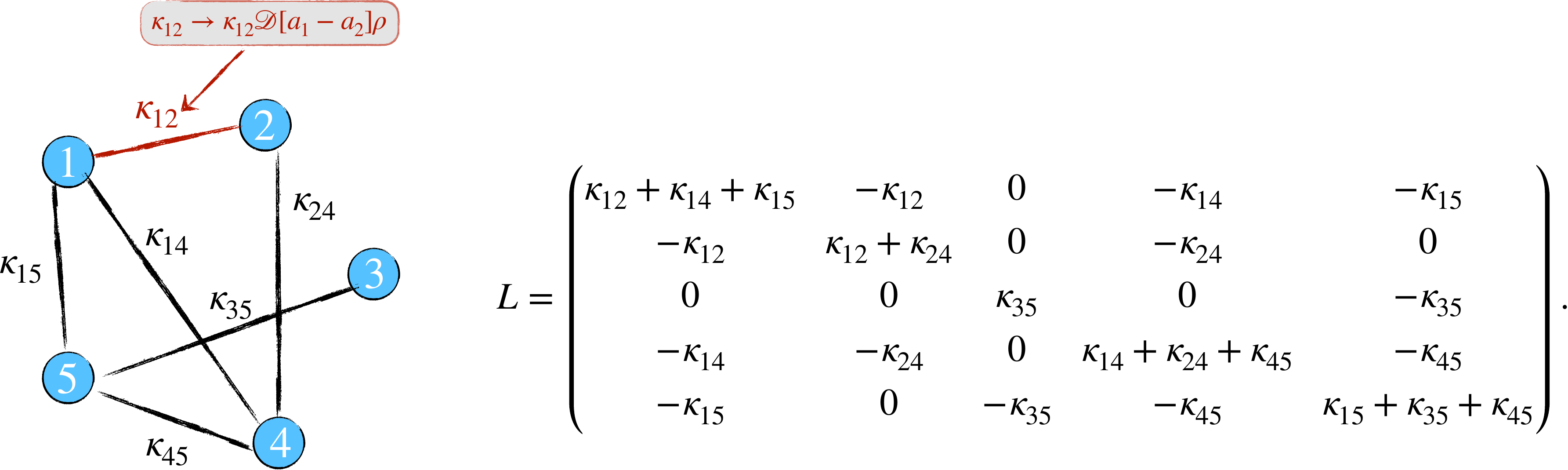}
    \caption{\label{Fig:graphL} Graph-Laplacian representation of the nonlocal dissipative coupling network.
    Each edge $(k,j)$ represents a dissipator $\kappa_{kj}\Ds[a_k-a_j]\rho$, while the corresponding graph Laplacian $L=D-K$ encodes the topology and coupling strengths of the network.}
\end{figure}
For an undirected network, $K$ is symmetric, with $\kappa_{kj}=\kappa_{jk}\geq 0$, and $K_{kk}=0$.
The corresponding degree matrix $D\in\mathbb{R}^{N\times N}$ is a diagonal matrix whose $k$th diagonal element gives the total dissipative coupling strength associated with oscillator $k$,
\begin{equation}
D_{kj}
=
\delta_{kj}\sum_{\ell\neq k}\kappa_{k\ell}.
\end{equation}
The graph Laplacian $L\in\mathbb{R}^{N\times N}$ is then defined as
\begin{equation}
L=D-K.
\end{equation}
To see its physical meaning, we introduce the operator vector $\mathbf{a}=(a_1,a_2,\ldots,a_N)^{\mathrm T}$.
The action of the graph Laplacian on $\mathbf{a}$ is given by
\begin{equation}
(L\mathbf{a})_k
=\sum_{j=1}^{N}L_{kj}a_j=
\sum_{j\neq k}\kappa_{kj}(a_k-a_j).
\label{Eq:Laplacian-action}
\end{equation}
Thus, the graph Laplacian naturally collects the relative differences $a_k-a_j$ over all dissipative links connected to oscillator $k$.
Using the definition of the graph Laplacian, the nonlocal dissipative coupling can be rewritten as
\begin{equation}
\sum_{k<j}\kappa_{kj}\Ds[a_k-a_j]\rho=
\sum_{k,j=1}^{N}L_{kj}
\left[
a_k\rho a_j^\dagger
-\frac{1}{2}
\left\{
a_j^\dagger a_k,\rho
\right\}
\right].
\label{Eq:Laplacian-dissipator}
\end{equation}

We illustrate the graph-Laplacian representation by considering a network of five oscillators, as shown in Fig.~\ref{Fig:graphL}, where each dissipative link is encoded by the corresponding elements of the Laplacian matrix.
For example, the dissipative coupling between oscillators $1$ and $2$, $\kappa_{12}\Ds[a_1-a_2]\rho$, contributes $-\kappa_{12}$ to the off-diagonal elements $L_{12}$ and $L_{21}$, and $\kappa_{12}$ to the diagonal elements $L_{11}$ and $L_{22}$.
The nonlocal dissipative channels involving oscillator $1$ are $\sum_{j\in\{2,4,5\}}\kappa_{1j}\Ds[a_1-a_j]\rho$.
Upon expansion, their local contribution associated with oscillator $1$ is $(\kappa_{12}+\kappa_{14}+\kappa_{15})\Ds[a_{1}]\rho$, which is precisely the diagonal element $L_{11}$.
The corresponding cross terms give $L_{1j}=L_{j1}=-\kappa_{1j}$ for $j\in \{2,4,5\}$.
The remaining rows of $L$ follow analogously.

\section{Mean-field approximation and classical limit cycles}

In this section, we derive the mean-field equations of motion and use them to study the classical limit-cycle dynamics and the associated relaxation behavior.
The mean-field treatment is justified in the large-amplitude regime, where quantum fluctuations and interoscillator correlations are negligible compared with the mean fields.
After introducing the classical complex amplitudes $\alpha_k=\langle a_k\rangle$ and factorizing the nonlinear moment as $\langle a_k^\dagger a_k a_k\rangle\simeq\langle a_k^\dagger\rangle\langle a_k\rangle^2=|\alpha_k|^2\alpha_k$, we obtain from the master equation~\eqref{Eq:ME} a closed set of equations for $\alpha_k$,
\begin{equation}
\dot{\alpha}_{k} = -i\omega\alpha_{k}+\frac{\gamma_{\uparrow}}{2}\alpha_{k}-\gamma_{\downarrow}|\alpha_{k}|^2\alpha_{k}-\sum_{j\neq k}\frac{\kappa_{kj}}{2}(\alpha_{k}-\alpha_{j}).
\end{equation}
With the help of the graph Laplacian, cf.
Eq.~\eqref{Eq:Laplacian-action}, the coupling term can be rewritten as $\sum_{j\neq k}\kappa_{kj}(\alpha_{k}-\alpha_{j})=(L\boldsymbol{\alpha})_{k}$, where $\boldsymbol{\alpha}=(\alpha_{1},\alpha_{2},\ldots,\alpha_{N})^{\mathrm T}$.
The mean-field equations thus take the compact form
\begin{equation}
\dot{\alpha}_{k} = -i\omega\alpha_{k}+\frac{\gamma_{\uparrow}}{2}\alpha_{k}-\gamma_{\downarrow}|\alpha_{k}|^2\alpha_{k}-\frac{1}{2}\sum_{j=1}^{N}L_{kj}\alpha_{j},
\label{Eq:MF-Laplacian}
\end{equation}
or, equivalently, in vector form
\begin{equation}
\dot{\boldsymbol{\alpha}} = \left(-i\omega+\frac{\gamma_{\uparrow}}{2}\right)\boldsymbol{\alpha}-\gamma_{\downarrow}
(\boldsymbol{\alpha}^{*}\circ\boldsymbol{\alpha}\circ\boldsymbol{\alpha})-\frac{1}{2}L\boldsymbol{\alpha},
\label{Eq:MF-Laplacian-vector}
\end{equation}
where $\circ$ denotes the elementwise (Hadamard) product, i.e., $(\boldsymbol{\alpha}^{*}\circ\boldsymbol{\alpha}\circ\boldsymbol{\alpha})_k=|\alpha_{k}|^2\alpha_{k}$.
In the following, we use these equations to analyze the classical limit-cycle dynamics of the network and, in particular, the relaxation of the system toward the synchronized limit cycle.

\blueit{The $N=2$ case.--} For two oscillators with coupling strength $\kappa\equiv\kappa_{12}$, the graph Laplacian reduces to $L=\kappa\bigl(\begin{smallmatrix}1&-1\\-1&1\end{smallmatrix}\bigr)$, and the mean-field equations~\eqref{Eq:MF-Laplacian} become
\begin{equation}
\begin{aligned}
\dot{\alpha}_{1} =& -i\omega\alpha_{1}+\frac{\gamma_{\uparrow}}{2}\alpha_{1}-\gamma_{\downarrow}|\alpha_{1}|^2\alpha_{1}-\frac{\kappa}{2}(\alpha_{1}-\alpha_{2}),   \\
\dot{\alpha}_{2} =& -i\omega\alpha_{2}+\frac{\gamma_{\uparrow}}{2}\alpha_{2}-\gamma_{\downarrow}|\alpha_{2}|^2\alpha_{2}-\frac{\kappa}{2}(\alpha_{2}-\alpha_{1}).
\end{aligned}
\label{Eq:MF-N2}
\end{equation}
We first consider the simplest initial condition, $\alpha_{1}(0)=\alpha_{2}(0)=\alpha_{0}$, in which the two oscillators start with identical amplitudes and phases.
The dissipative coupling term then vanishes identically, and the dynamics of each oscillator reduces to that of a single vdP oscillator relaxing toward its limit cycle.
To show this explicitly, we write the common amplitude as $\alpha_{1}=\alpha_{2}\equiv\alpha=r e^{i\phi}$ with real $r\geq0$ and $\phi$, so that the mean-field equation separates into
\begin{equation}
\dot{r}=\frac{\gamma_{\uparrow}}{2}\,r-\gamma_{\downarrow}r^{3},
\qquad
\dot{\phi}=-\omega .
\label{Eq:MF-polar}
\end{equation}
The phase thus simply rotates at the bare frequency, while the radial dynamics admits two fixed points, $r=0$ and
\begin{equation}
r_{\text{LC}}=\sqrt{\frac{\gamma_{\uparrow}}{2\gamma_{\downarrow}}}.
\label{Eq:rlc}
\end{equation}
Linearizing Eq.~\eqref{Eq:MF-polar} around them shows that the origin is unstable, with radial growth rate $\gamma_{\uparrow}/2>0$, whereas $r_{\text{LC}}$ is asymptotically stable, with decay rate $-\gamma_{\uparrow}<0$ of radial perturbations.
In fact, the radial equation can be solved exactly: for the intensity $n(t)=r^{2}(t)$, Eq.~\eqref{Eq:MF-polar} yields the logistic equation $\dot{n}=\gamma_{\uparrow}n-2\gamma_{\downarrow}n^{2}$, whose solution reads
\begin{equation}
n(t)=\frac{r_{\text{LC}}^{2}}{1+\bigl(r_{\text{LC}}^{2}/n_{0}-1\bigr)e^{-\gamma_{\uparrow}t}},
\qquad
n_{0}=|\alpha_{0}|^{2}.
\label{Eq:MF-logistic}
\end{equation}
Hence, for any $\alpha_{0}\neq0$, $n(t)\to r_{\text{LC}}^{2}$ as $t\to\infty$: the classical trajectory approaches the limit cycle $\alpha(t)=r_{\text{LC}}\,e^{-i\omega t+i\phi_{0}}$, and the radial relaxation toward it is asymptotically exponential with rate $\gamma_{\uparrow}$.

For a generic initial condition, $\alpha_{1}(0)\neq\alpha_{2}(0)$, we define the difference $\delta\alpha=\alpha_{1}-\alpha_{2}$, which evolves according to
\begin{equation}
 \dot{\delta\alpha} = \dot{\alpha}_{1}-\dot{\alpha}_{2} = \left(-i\omega+\frac{\gamma_{\uparrow}}{2}-\kappa\right)\delta\alpha-\gamma_{\downarrow}\left(|\alpha_{1}|^{2}\alpha_{1}-|\alpha_{2}|^{2}\alpha_{2}\right).
\end{equation}
By introducing the Lyapunov function $V=\frac{1}{2}|\delta\alpha|^{2}\geq 0$, we obtain
\begin{equation}
    \dot{V}=(\gamma_{\uparrow}-2\kappa)V-\frac{\gamma_{\downarrow}}{2}\left[(|\alpha_1|^{2}+|\alpha_2|^{2})|\delta\alpha|^2+(|\alpha_1|^{2}-|\alpha_2|^{2})^2\right].
\label{Eq:Lyapunov}
\end{equation}
Both terms inside the square brackets are nonnegative and vanish only on the synchronized manifold $\alpha_{1}=\alpha_{2}$, so that
\begin{equation}
\dot{V}\leq(\gamma_{\uparrow}-2\kappa)V
\quad\Longrightarrow\quad
V(t)\leq V(0)\,e^{-(2\kappa-\gamma_{\uparrow})t}.
\label{Eq:Lyapunov-bound}
\end{equation}
Hence, for strong coupling, $\kappa>\gamma_{\uparrow}/2$, the synchronization manifold is globally attractive, and any initial condition approaches it exponentially with a guaranteed convergence rate of at least $\kappa-\frac{\gamma_{\uparrow}}{2}$.
It should be emphasized that this bound is only a sufficient lower bound on the convergence rate and is generally not tight.
This is because the Lyapunov analysis discards the additional negative contribution of the nonlinear damping term.
Therefore, synchronization may occur faster than predicted by this bound, and the condition $\kappa>\gamma_{\uparrow}/2$ should not be interpreted as the actual synchronization threshold.

\blueit{The $N>2$ case.--} For a general network of $N$ oscillators, the relaxation dynamics is transparent in the eigenbasis of the graph Laplacian.
We first introduce the average amplitude $\bar{\alpha}=\frac{1}{N}\sum_{k=1}^{N}\alpha_{k}$ and the synchronization error $e_{k}=\alpha_{k}-\bar{\alpha}$, which measures the deviation of each oscillator from the mean.
Then we define the Lyapunov function $V=\frac{1}{2}\boldsymbol{e}^{\dagger}\boldsymbol{e}=\frac{1}{2}\sum_{k=1}^{N}|e_{k}|^{2}\geq 0$, where $\boldsymbol{e}=(e_{1},e_{2},\ldots,e_{N})^{\mathrm T}$.
The function $V$ vanishes only on the synchronized manifold $\alpha_{1}=\alpha_{2}=\cdots=\alpha_{N}$.

From Eq.~\eqref{Eq:MF-Laplacian-vector}, we obtain $\dot{V}=\mathrm{Re}\sum_{k}e_{k}^{*}\dot{e}_{k}=\mathrm{Re}\sum_{k}e_{k}^{*}\dot{\alpha}_{k}$ with $(\dot{\boldsymbol{e}}^{\dagger}\boldsymbol{e})^{*}=\boldsymbol{e}^{\dagger}\dot{\boldsymbol{e}}$, where the second equality follows from $\sum_{k}e_{k}=0$, and hence
\begin{equation}
    \dot{V}=\gamma_{\uparrow}V-\gamma_{\downarrow}\,\mathrm{Re}\sum_{k}e_{k}^{*}|\alpha_{k}|^2\alpha_{k} -\frac{1}{2}\boldsymbol{e}^{\dagger}L\boldsymbol{e}.
\label{Eq:Lyapunov-N}
\end{equation}

To bound the last term, we recall the spectral properties of the graph Laplacian for an undirected weighted network.
First, $L$ is real symmetric, $L=L^{\mathrm T}$, so its eigenvalues are real and its eigenvectors can be chosen orthonormal.
Second, each row sums to zero, $\sum_{j}L_{kj}=0$, which implies that the uniform vector $\boldsymbol{1}=(1,1,\ldots,1)^{\mathrm T}$ is a zero mode, $L\boldsymbol{1}=0$.
Finally, for any $\boldsymbol{x}\in\mathbb{C}^{N}$,
\begin{equation}
\boldsymbol{x}^{\dagger}L\boldsymbol{x}=\frac{1}{2}\sum_{k,j}\kappa_{kj}|x_k-x_j|^2\geq 0,
\label{Eq:L-quadratic}
\end{equation}
so $L$ is positive semidefinite.
The spectrum thus consists of $0=\lambda_{1}\leq\lambda_{2}\leq\cdots\leq\lambda_{N}$ with orthonormal eigenvectors $u_{\mu}$; for a connected network the zero eigenvalue is simple, i.e., $\lambda_{2}>0$.
Crucially, the synchronization error has no component along the zero mode, $\boldsymbol{e}^{\dagger}\boldsymbol{1}=\sum_{k}e_{k}=0$, so the Rayleigh variational principle yields
\begin{equation}
\boldsymbol{e}^{\dagger}L\boldsymbol{e}\geq\lambda_{2}\,\boldsymbol{e}^{\dagger}\boldsymbol{e}=2\lambda_{2}V.
\label{Eq:L-Rayleigh}
\end{equation}
The nonlinear term in Eq.~\eqref{Eq:Lyapunov-N} is nonnegative as well: using again $\sum_{k}e_{k}=0$, it can be recast as a sum over pairs,
\begin{equation}
\mathrm{Re}\sum_{k}e_{k}^{*}|\alpha_{k}|^{2}\alpha_{k}
=\frac{1}{4N}\sum_{k,j}\Bigl[\bigl(|\alpha_{k}|^{2}+|\alpha_{j}|^{2}\bigr)|\alpha_{k}-\alpha_{j}|^{2}+\bigl(|\alpha_{k}|^{2}-|\alpha_{j}|^{2}\bigr)^{2}\Bigr]\geq0 .
\label{Eq:nonlinear-positive}
\end{equation}
Combining Eqs.~\eqref{Eq:L-Rayleigh} and~\eqref{Eq:nonlinear-positive} with Eq.~\eqref{Eq:Lyapunov-N}, we arrive at
\begin{equation}
\dot{V}\leq-(\lambda_{2}-\gamma_{\uparrow})V
\quad\Longrightarrow\quad
V(t)\leq V(0)\,e^{-(\lambda_{2}-\gamma_{\uparrow})t}.
\label{Eq:Lyapunov-bound-N}
\end{equation}
Hence, for any connected network with algebraic connectivity $\lambda_{2}>\gamma_{\uparrow}$, the synchronized manifold is globally attractive, and the synchronization error decays exponentially at a guaranteed rate of at least $(\lambda_{2}-\gamma_{\uparrow})/2$.
For $N=2$, where $\lambda_{2}=2\kappa$, this reduces to the bound obtained above.

\section{Many-body synchronization measure}

In this section, we generalize the synchronization measure introduced by \citet{Mari2013} for the two-vdP oscillator case to networks of $N$ vdP oscillators.
For two oscillators, the Mari's complete synchronization measure is defined as
\begin{equation}
    \Ss_{c}(t)=\frac{1}{\langle q_{-}(t)^{2}+p_{-}(t)^{2}\rangle},
\end{equation}
where $q_{-}(t)=(q_{1}(t)-q_{2}(t))/\sqrt{2}$ and $p_{-}(t)=(p_{1}(t)-p_{2}(t))/\sqrt{2}$ denote the relative quadratures between the two oscillators.
Here, the quadrature operators of the $k$th oscillator are defined as $q_{k}=(a_{k}+a_{k}^{\dagger})/\sqrt{2}$ and $p_{k}=i(a_{k}^{\dagger}-a_{k})/\sqrt{2}$.
The denominator can be decomposed as
\begin{equation}
\left\langle q_-^2+p_-^2\right\rangle
=
\langle q_-\rangle^2+\langle p_-\rangle^2
+(\Delta q_-)^2+(\Delta p_-)^2,
\end{equation}
showing that $\Ss_c$ accounts for both the mismatch between the mean quadratures and their quantum fluctuations.
Since $[q_-,p_-]=i$, the Heisenberg uncertainty relation implies $(\Delta q_-)^2+(\Delta p_-)^2 \geq 2\Delta q_-\Delta p_-\geq1$.
Consequently, $0<\Ss_c(t)\leq1$.
The upper bound is reached when the mean quadrature differences vanish and the relative mode has minimum, balanced quantum fluctuations.
Thus, even when the mean trajectories are perfectly synchronized, quantum fluctuations remain explicitly included in $\Ss_c$.

Motivated by this construction, we generalize the complete synchronization measure to a network of $N$ oscillators using the eigenmodes of the graph Laplacian.
To this end, we first identify the synchronized mode and the modes transverse to it.

Introducing the quadrature vectors $\boldsymbol{q}=(q_1,q_2,\ldots,q_N)^{\mathrm T}$, $\boldsymbol{p}=(p_1,p_2,\ldots,p_N)^{\mathrm T}$, we define complete synchronization at the level of the first moments, i.e., all oscillators have identical mean quadratures,
\begin{equation}
\langle q_1\rangle=\cdots=\langle q_N\rangle,
\qquad
\langle p_1\rangle=\cdots=\langle p_N\rangle.
\end{equation}
Consequently, the vectors $\langle\boldsymbol q\rangle$ and $\langle\boldsymbol p\rangle$ lie along the uniform direction
\begin{equation}
\boldsymbol u_1
=\frac{1}{\sqrt N}(1,1,\ldots,1)^{\mathrm T}.
\end{equation}
For a connected undirected network with a real symmetric graph Laplacian $L$, this direction is the unique normalized zero mode,
\begin{equation}
L\boldsymbol u_1=0.
\end{equation}
The Laplacian eigenvectors $\{\boldsymbol u_\mu\}_{\mu=1}^{N}$ then form an orthonormal basis of the oscillator space.
Thus, the uniform mode $\boldsymbol u_1$ defines the collective synchronized mode, while the remaining eigenvectors $\boldsymbol u_\mu$, $\mu=2,\ldots,N$, form a convenient orthonormal basis of the transverse subspace.
The corresponding projections
\begin{equation}
Q_\mu=\boldsymbol u_\mu^{\mathrm T}\boldsymbol q,
\qquad
P_\mu=\boldsymbol u_\mu^{\mathrm T}\boldsymbol p
\end{equation}
describe deviations from the completely synchronized manifold.
We expand these quadrature vectors in the orthonormal eigenbasis of the graph Laplacian.
The quadratures associated with the $\mu$th Laplacian mode are defined as
\begin{equation}
Q_{\mu}(t)
=\boldsymbol{u}_{\mu}^{\mathrm T}\boldsymbol{q}(t)
=\sum_{k=1}^{N}u_{\mu k}q_k(t),
\qquad
P_{\mu}(t)
=\boldsymbol{u}_{\mu}^{\mathrm T}\boldsymbol{p}(t)
=\sum_{k=1}^{N}u_{\mu k}p_k(t),
\label{Eq:mode-quadratures}
\end{equation}
where $u_{\mu k}$ is the $k$th component of $\boldsymbol{u}_{\mu}$.
The pair $(Q_1,P_1)$ describes the collective quadratures along the synchronized manifold, whereas $(Q_{\mu},P_{\mu})$ with $\mu=2,\ldots,N$ describe the transverse quadratures.
Accordingly, we quantify the departure from complete synchronization by the mean-square fluctuations of all transverse quadratures and define
\begin{equation}
\Ss_{\perp}(t)=
\left[
\frac{1}{N-1}
\sum_{\mu=2}^{N}
\left\langle Q_{\mu}(t)^{2}+P_{\mu}(t)^{2}\right\rangle
\right]^{-1}.
\label{Eq:many-body-sync}
\end{equation}
The uniform mode $\mu=1$ is excluded because it describes the collective motion along the synchronized manifold and does not measure deviations from synchronization.
The factor $1/(N-1)$ averages over the transverse modes, making the normalization independent of the system size.
A larger $\Ss_{\perp}$ therefore indicates stronger synchronization.
In the following, we discuss the key properties of this measure.

\blueit{Canonical structure and quantum bound.--} Since the Laplacian eigenvectors are orthonormal and
$[q_k,p_j]=i\delta_{kj}$, the mode transformation preserves the canonical structure,
\begin{equation}
[Q_\mu,P_\nu]
=i\sum_{k}u_{\mu k}u_{\nu k}
=i\delta_{\mu\nu}.
\end{equation}
The uncertainty relation therefore gives, for each transverse mode,
\begin{equation}
\left\langle Q_\mu^2+P_\mu^2\right\rangle
\geq(\Delta Q_\mu)^2+(\Delta P_\mu)^2
\geq1.
\end{equation}
Consequently, the many-body synchronization measure is bounded as
\begin{equation}
0<\Ss_{\perp}(t)\leq1.
\end{equation}

\blueit{Independence of the transverse basis.--}
The choice of the transverse basis is not unique: any two orthonormal bases orthogonal to the synchronized mode $\boldsymbol{u}_1$ are related by an orthogonal transformation $O\in O(N-1)$.
In particular, $\widetilde{\boldsymbol{u}}_{\mu}=\sum_{\nu=2}^{N}O_{\mu\nu}\boldsymbol{u}_{\nu}$, with $ O^{\mathrm T}O=I$.
The corresponding quadratures transform as
\begin{equation}
\widetilde{Q}_{\mu}
=
\sum_{\nu=2}^{N}O_{\mu\nu}Q_{\nu},
\qquad
\widetilde{P}_{\mu}
=
\sum_{\nu=2}^{N}O_{\mu\nu}P_{\nu}.
\end{equation}
Orthogonality of $O$ then implies
\begin{equation}
\sum_{\mu=2}^{N}
\left(\widetilde{Q}_{\mu}^{2}+\widetilde{P}_{\mu}^{2}\right)
=
\sum_{\mu=2}^{N}
\left(Q_{\mu}^{2}+P_{\mu}^{2}\right).
\end{equation}
Thus, $\Ss_{\perp}$ is independent of the particular transverse basis.
The transverse Laplacian eigenvectors provide a convenient, but not essential, choice of basis.

Equivalently, the basis independence can be made explicit by introducing the orthogonal projector onto the transverse subspace.
The transverse projector is
\begin{equation}
\Pi_{\perp}
=
\sum_{\mu=2}^{N}
\boldsymbol{u}_{\mu}\boldsymbol{u}_{\mu}^{\mathrm T}
=
I-\boldsymbol{u}_1\boldsymbol{u}_1^{\mathrm T}
=
I-\frac{1}{N}\boldsymbol{1}\boldsymbol{1}^{\mathrm T}.
\label{Eq:transverse-projector}
\end{equation}
Indeed, using $Q_\mu=\boldsymbol{u}_{\mu}^{\mathrm T}\boldsymbol{q}$, we find
\begin{equation}
\begin{aligned}
\boldsymbol{q}^{\mathrm T}\Pi_{\perp}\boldsymbol{q}
&=
\boldsymbol{q}^{\mathrm T}
\left(
\sum_{\mu=2}^{N}
\boldsymbol{u}_{\mu}\boldsymbol{u}_{\mu}^{\mathrm T}
\right)
\boldsymbol{q} \\
&=
\sum_{\mu=2}^{N}
\left(
\boldsymbol{u}_{\mu}^{\mathrm T}\boldsymbol{q}
\right)^2
=
\sum_{\mu=2}^{N}Q_\mu^2.
\end{aligned}
\end{equation}
Similarly,
\begin{equation}
\boldsymbol{p}^{\mathrm T}\Pi_{\perp}\boldsymbol{p}
=
\sum_{\mu=2}^{N}P_\mu^2.
\end{equation}
Therefore, the synchronization measure can be written in projector form as
\begin{equation}
\Ss_{\perp}(t)
=
\left[
\frac{1}{N-1}
\left\langle
\boldsymbol{q}^{\mathrm T}(t)\Pi_{\perp}\boldsymbol{q}(t)
+
\boldsymbol{p}^{\mathrm T}(t)\Pi_{\perp}\boldsymbol{p}(t)
\right\rangle
\right]^{-1}.
\label{Eq:projector-form}
\end{equation}
Since $\Pi_{\perp}$ depends only on the uniform synchronized mode $\boldsymbol{u}_1$, this expression is independent of the particular orthonormal basis chosen within the transverse subspace.

\blueit{Pairwise representation and exact reduction to Mari's measure.--}
The projector form in Eq.~\eqref{Eq:projector-form} admits an equivalent pairwise representation.
Using $\Pi_{\perp}=I-\boldsymbol{1}\boldsymbol{1}^{\mathrm T}/N$, we obtain
\begin{equation}
\begin{aligned}
\boldsymbol{q}^{\mathrm T}\Pi_{\perp}\boldsymbol{q}
&=
\sum_{k=1}^{N}q_k^2
-\frac{1}{N}\left(\sum_{k=1}^{N}q_k\right)^2 \\
&=
\frac{1}{N}\sum_{k<j}(q_k-q_j)^2.
\end{aligned}
\end{equation}
Similarly,
\begin{equation}
\boldsymbol{p}^{\mathrm T}\Pi_{\perp}\boldsymbol{p}
=
\frac{1}{N}\sum_{k<j}(p_k-p_j)^2.
\end{equation}
It follows that
\begin{equation}
\Ss_{\perp}(t)
=
\left[
\frac{1}{N(N-1)}
\sum_{k<j}
\left\langle
(q_k(t)-q_j(t))^2+(p_k(t)-p_j(t))^2
\right\rangle
\right]^{-1}.
\label{Eq:pairwise-sync}
\end{equation}
For $N=2$, Eq.~\eqref{Eq:pairwise-sync} reduces to
\begin{equation}
\begin{aligned}
\Ss_{\perp}(t)
&=
\left[
\frac{1}{2}
\left\langle
(q_1(t)-q_2(t))^2+(p_1(t)-p_2(t))^2
\right\rangle
\right]^{-1} \\
&=
\frac{1}{\left\langle q_-(t)^2+p_-(t)^2\right\rangle}
=
\Ss_c(t).
\end{aligned}
\end{equation}
Thus, for two oscillators, the many-body measure reduces exactly to the complete synchronization measure introduced by Mari et al.~\cite{Mari2013}.

\section{Truncated-Wigner approximation and exact benchmark}

To investigate the synchronization dynamics of large oscillator networks, we employ the truncated-Wigner approximation (TWA)~\cite{Polkovnikov2010,Bond2024}, since the Hilbert-space dimension grows exponentially with the number of oscillators, ruling out an exact solution of the master equation for large networks.
The TWA maps the quantum master equation onto stochastic dynamics of complex phase-space variables while retaining the leading quantum fluctuations around the semiclassical evolution.
In the following, we outline the derivation of the corresponding stochastic equations for the dissipatively coupled quantum vdP oscillator network and explain how the synchronization measure is evaluated from stochastic trajectories.
We then benchmark the approximation against the exact master-equation dynamics for two oscillators.

The starting point is the Wigner function of the $N$-oscillator network.
Introducing the vectors of phase-space variables $\boldsymbol{\alpha}=(\alpha_{1},\ldots,\alpha_{N})^{\mathrm T}$ and $\boldsymbol{\lambda}=(\lambda_{1},\ldots,\lambda_{N})^{\mathrm T}$, the Wigner function associated with the density operator $\rho$ is defined as~\cite{Walls2008}
\begin{equation}
   W(\boldsymbol{\alpha},\boldsymbol{\alpha}^{*})=\frac{1}{\pi^{2N}}\int d^{2}\boldsymbol{\lambda}\, e^{\boldsymbol{\lambda}^{*}\cdot\boldsymbol{\alpha}-\boldsymbol{\lambda}\cdot\boldsymbol{\alpha}^{*}}\,\mathrm{Tr}\!\left[\rho\, e^{\boldsymbol{\lambda}\cdot\mathbf{a}^{\dagger}-\boldsymbol{\lambda}^{*}\cdot\mathbf{a}}\right],
\end{equation}
with the operator vector $\mathbf{a}=(a_{1},\ldots,a_{N})^{\mathrm T}$ and $d^{2}\boldsymbol{\lambda}=\prod_{k=1}^{N}d^{2}\lambda_{k}$.
For each mode $k$, the action of the creation and annihilation operators on $\rho$ translates into differential operations on $W(\boldsymbol{\alpha},\boldsymbol{\alpha}^{*})$ according to the correspondence rules
\begin{equation}
\begin{aligned}
    a_{k}\rho&\rightarrow\left(\alpha_{k}+\frac{1}{2}\frac{\partial}{\partial\alpha_{k}^{*}}\right)W(\boldsymbol{\alpha},\boldsymbol{\alpha}^{*}),\quad
    a^{\dagger}_{k}\rho\rightarrow\left(\alpha_{k}^{*}-\frac{1}{2}\frac{\partial}{\partial\alpha_{k}}\right)W(\boldsymbol{\alpha},\boldsymbol{\alpha}^{*}), \\
    \rho a_{k}&\rightarrow\left(\alpha_{k}-\frac{1}{2}\frac{\partial}{\partial\alpha_{k}^{*}}\right)W(\boldsymbol{\alpha},\boldsymbol{\alpha}^{*}),\quad
    \rho a^{\dagger}_{k}\rightarrow\left(\alpha_{k}^{*}+\frac{1}{2}\frac{\partial}{\partial\alpha_{k}}\right)W(\boldsymbol{\alpha},\boldsymbol{\alpha}^{*}).
\end{aligned}
\end{equation}
We then apply these rules term by term to the master equation~\eqref{Eq:ME}, converting it into a partial differential equation for the Wigner function.
Collecting the resulting terms into divergences gives its Kramers--Moyal form
\begin{equation}
\begin{aligned}
\frac{\partial W}{\partial t}={}&i\omega\sum_{k=1}^{N}\left[\frac{\partial}{\partial \alpha_{k}}(\alpha_{k}W)-\frac{\partial}{\partial \alpha_{k}^{*}}(\alpha_{k}^{*}W) \right]-\frac{\gamma_{\uparrow}}{2}\sum_{k=1}^{N}\left[\frac{\partial}{\partial \alpha_{k}}(\alpha_{k}W)+\frac{\partial}{\partial \alpha_{k}^{*}}(\alpha_{k}^{*}W)-\frac{\partial^2W}{\partial \alpha_{k}\,\partial \alpha_{k}^{*}} \right]\\
&+\gamma_{\downarrow}\sum_{k=1}^{N}\Bigg\{
\frac{\partial}{\partial\alpha_k}\left[(|\alpha_k|^2-1)\alpha_kW\right]
+\frac{\partial}{\partial\alpha_k^*}\left[(|\alpha_k|^2-1)\alpha_k^*W\right]\\
&+\frac{\partial^2}{\partial\alpha_k\,\partial\alpha_k^*}
\left[(2|\alpha_k|^2-1)W\right]
+\frac{1}{4}\frac{\partial^3}{\partial\alpha_k^2\,\partial\alpha_k^*}(\alpha_kW)
+\frac{1}{4}\frac{\partial^3}{\partial\alpha_k\,\partial(\alpha_k^*)^2}(\alpha_k^*W)\Bigg\}\\
&+\sum_{k<j}\frac{\kappa_{kj}}{2}\Big\{\left[\frac{\partial}{\partial \alpha_{k}}-\frac{\partial}{\partial \alpha_{j}}\right]\left[(\alpha_{k}-\alpha_{j})W\right]+\left[\frac{\partial}{\partial \alpha_{k}^{*}}-\frac{\partial}{\partial \alpha_{j}^{*}}\right]\left[(\alpha_{k}^{*}-\alpha_{j}^{*})W\right]+\left[\frac{\partial}{\partial \alpha_{k}}-\frac{\partial}{\partial \alpha_{j}}\right]\left[\frac{\partial}{\partial \alpha_{k}^{*}}-\frac{\partial}{\partial \alpha_{j}^{*}}\right]W\Big\}.
\end{aligned}
\label{Eq:W-KM}
\end{equation}
For a compact representation of Eq.~\eqref{Eq:W-KM}, let $n_k=|\alpha_k|^2$ and define the drift vector
\begin{equation}
A_k=\left[-i\omega+\frac{\gamma_{\uparrow}}{2}
-\gamma_{\downarrow}(n_k-1)\right]\alpha_k
-\frac{1}{2}\sum_{j=1}^{N}L_{kj}\alpha_j .
\label{Eq:KM-drift}
\end{equation}
The exact Wigner equation can then be written as
\begin{equation}
\frac{\partial W}{\partial t}=-\sum_k\left[
\frac{\partial}{\partial\alpha_k}(A_kW)
+\frac{\partial}{\partial\alpha_k^*}(A_k^*W)\right]+\sum_{k,j}\frac{\partial^2}{\partial\alpha_k\partial\alpha_j^*}
\left(C_{kj}^{(2)}W\right)+\frac{\gamma_{\downarrow}}{4}\sum_k\left[
\frac{\partial^3}{\partial\alpha_k^2\partial\alpha_k^*}(\alpha_kW)
+\frac{\partial^3}{\partial\alpha_k\partial(\alpha_k^*)^2}(\alpha_k^*W)
\right],
\label{Eq:KM-exact}
\end{equation}
with the second-order Kramers--Moyal coefficient
\begin{equation}
C_{kj}^{(2)}=
\left[\frac{\gamma_{\uparrow}}{2}
+\gamma_{\downarrow}(2n_k-1)\right]\delta_{kj}
+\frac{1}{2}L_{kj}.
\label{Eq:KM-second}
\end{equation}
The third-order terms in Eq.~\eqref{Eq:KM-exact} originate from the nonlinear two-photon loss; all remaining contributions already have Fokker--Planck form.
The truncated-Wigner approximation consists in discarding these third-order terms while retaining the full drift~\eqref{Eq:KM-drift} and diffusion~\eqref{Eq:KM-second}, which yields the Fokker--Planck equation
\begin{equation}
\frac{\partial W}{\partial t}=-\sum_k\left[
\frac{\partial}{\partial\alpha_k}(A_kW)
+\frac{\partial}{\partial\alpha_k^*}(A_k^*W)\right]
+\sum_{k,j}\frac{\partial^2}{\partial\alpha_k\partial\alpha_j^*}
\left(C_{kj}^{(2)}W\right).
\label{Eq:TWA-FPE}
\end{equation}
Since the jump operators of the nonlocal dissipator are linear, its contribution---the Laplacian term in the drift~\eqref{Eq:KM-drift} and the $L_{kj}/2$ part of the diffusion~\eqref{Eq:KM-second}---is exact and generates no terms beyond second order.

Equivalently, the Fokker--Planck equation~\eqref{Eq:TWA-FPE} is sampled by the It\^o equations
\begin{equation}
d\alpha_k=A_k\,dt
+\sqrt{\frac{\gamma_{\uparrow}}{2}
+\gamma_{\downarrow}(2|\alpha_k|^2-1)}\,d\zeta_k
+\sum_{e=(r,s),\,r<s}\sqrt{\frac{\kappa_{rs}}{2}}
(\delta_{kr}-\delta_{ks})\,d\zeta_e,
\label{Eq:TWA-SDE}
\end{equation}
where $d\zeta_k$ and $d\zeta_e$ are mutually independent complex Wiener increments associated with oscillator $k$ and with the dissipative edge $e=(r,s)$, respectively, satisfying $\langle d\zeta_a d\zeta_b^*\rangle=\delta_{ab}dt$ and $\langle d\zeta_a d\zeta_b\rangle=0$.
In the last term, the sum runs over all dissipative edges of the network, and $\kappa_{rs}$ is the dissipative coupling strength of the edge $(r,s)$ entering Eq.~\eqref{Eq:ME}, i.e., $\kappa_{rs}\equiv\kappa_{kj}$ for $(r,s)=(k,j)$; the dummy indices are renamed only to avoid confusion with the oscillator index $k$.
The factor $(\delta_{kr}-\delta_{ks})$ adds opposite noise increments to the two endpoints of the edge; the edge noise therefore acts only on the relative mode of the pair, mirroring the structure of the underlying jump operator $a_r-a_s$.
Summing these edge contributions over the network reproduces the Laplacian part of the diffusion matrix, such that $\langle d\alpha_k d\alpha_j^{*}\rangle_{\mathrm{noise}} =C_{kj}^{(2)}dt$.

For illustration, consider two oscillators with $\kappa\equiv\kappa_{12}$.
In this case, $L=\kappa\left(\begin{smallmatrix}1&-1\\-1&1\end{smallmatrix}\right)$, and Eq.~\eqref{Eq:TWA-SDE} becomes
\begin{equation}
\begin{aligned}
d\alpha_1={}&\left\{\left[-i\omega+\frac{\gamma_{\uparrow}}{2}
-\gamma_{\downarrow}(|\alpha_1|^2-1)\right]\alpha_1
-\frac{\kappa}{2}(\alpha_1-\alpha_2)\right\}dt+\sqrt{\frac{\gamma_{\uparrow}}{2}
+\gamma_{\downarrow}(2|\alpha_1|^2-1)}\,d\zeta_1
+\sqrt{\frac{\kappa}{2}}\,d\zeta_{12},\\
d\alpha_2={}&\left\{\left[-i\omega+\frac{\gamma_{\uparrow}}{2}
-\gamma_{\downarrow}(|\alpha_2|^2-1)\right]\alpha_2
-\frac{\kappa}{2}(\alpha_2-\alpha_1)\right\}dt+\sqrt{\frac{\gamma_{\uparrow}}{2}
+\gamma_{\downarrow}(2|\alpha_2|^2-1)}\,d\zeta_2
-\sqrt{\frac{\kappa}{2}}\,d\zeta_{12}.
\end{aligned}
\label{Eq:TWA-SDE-N2}
\end{equation}
\begin{figure}[!b]
    \centering
    \includegraphics[width=0.65\textwidth]{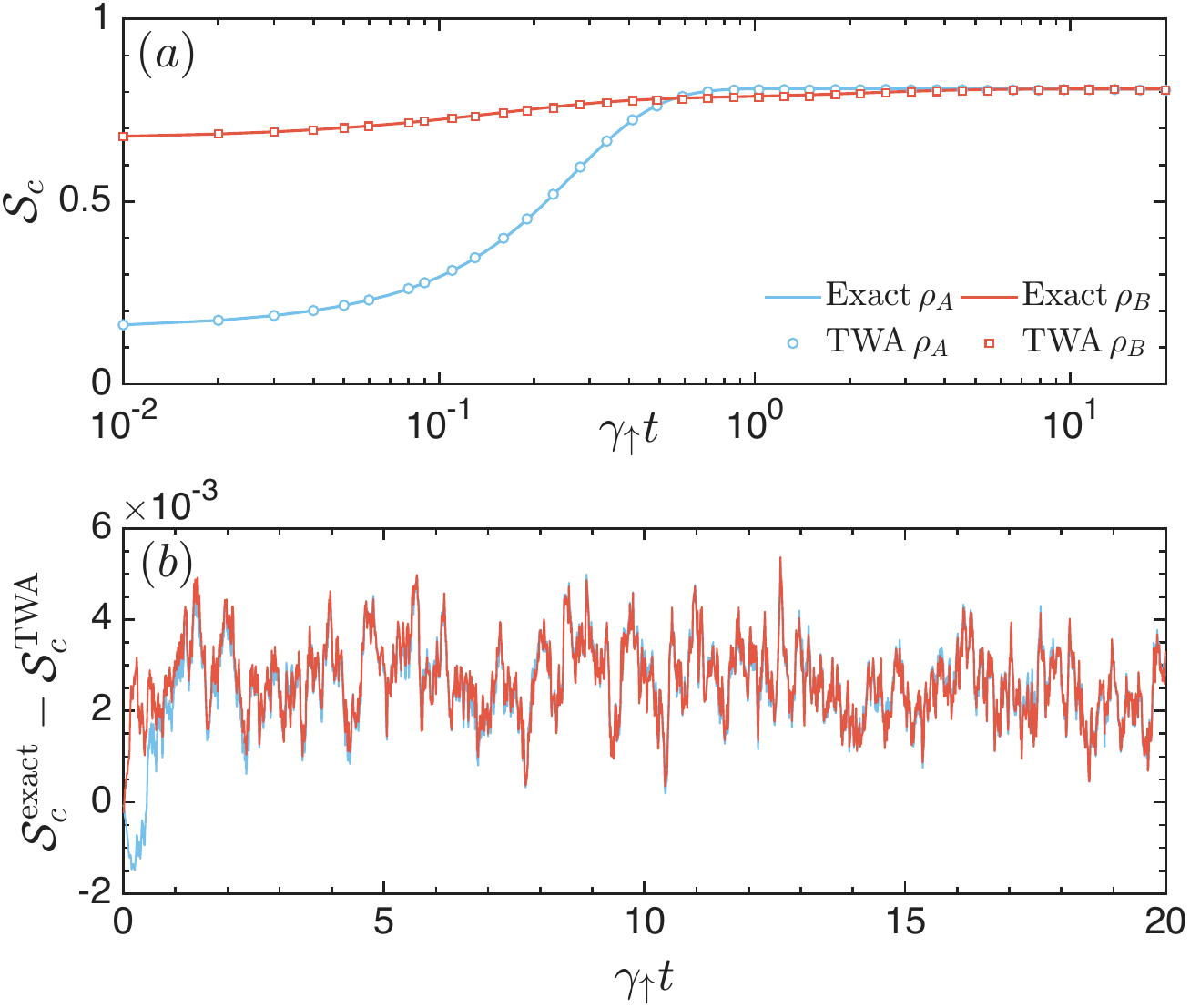}
    \caption{\label{Fig:N2TWA} Benchmark of the TWA against the exact master-equation dynamics for two dissipatively coupled vdP oscillators.
    (a) Synchronization measure $\Ss_{c}(t)$ for the initial states $\rho_{A}$ (blue) and $\rho_{B}$ (red): exact results (solid lines) versus TWA results (circles and squares).
    The TWA faithfully reproduces the exact dynamics, including the crossing of the two curves that signals the synchronization Mpemba effect.
    (b) Deviation $\Ss_{c}^{\rm exact}-\Ss_{c}^{\rm TWA}$ between the exact and TWA results, which remains at the level of $10^{-3}$ for both initial states.
    The parameters and the initial states are the same as in the main text: $\omega=\gamma_{\uparrow}$, $\gamma_{\downarrow}=0.1\gamma_{\uparrow}$, and $\kappa=4\gamma_{\uparrow}$.
    The TWA results are obtained by averaging $10^{6}$ stochastic trajectories of Eq.~\eqref{Eq:TWA-SDE-N2} integrated with the time step $\gamma_{\uparrow}dt=0.0025$.
}
\end{figure}
Here $d\zeta_1$, $d\zeta_2$, and $d\zeta_{12}$ are mutually independent, and contracting the noise terms gives
\begin{equation}
\left\langle d\alpha_1d\alpha_2^*\right\rangle_{\mathrm{noise}}
=-\frac{\kappa}{2}\,dt,
\qquad
\left\langle d(\alpha_1-\alpha_2)d(\alpha_1-\alpha_2)^*\right\rangle_{\mathrm{edge\text{-}noise}}
=2\kappa\,dt.
\label{Eq:TWA-N2-covariance}
\end{equation}
The cross correlator comes solely from the edge noise, since the site noises are independent; the opposite signs with which the edge noise enters $\alpha_1$ and $\alpha_2$ enhance its contribution to the relative coordinate to $2\kappa\,dt$.
Thus both the deterministic coupling and the edge noise act only on the relative mode of the pair: the in-phase combination $\alpha_1+\alpha_2$ feels neither the drift $-\frac{\kappa}{2}(\alpha_1-\alpha_2)$ nor the edge noise, while the relative mode is damped at the rate $\kappa$.
This is the stochastic counterpart of the dark-mode structure of the underlying dissipator $\kappa\Ds[a_1-a_2]\rho$.

To benchmark the TWA against the exact dynamics, we evaluate the synchronization measure from stochastic trajectories.
For two oscillators, $\Ss_{c}$ involves only the expectation value $\langle q_{-}^{2}+p_{-}^{2}\rangle$, which is symmetrically ordered in the creation and annihilation operators.
Within the TWA, such Weyl-ordered expectation values are evaluated as classical moments of the stochastic phase-space variables, so that
\begin{equation}
\left\langle q_{-}(t)^{2}+p_{-}(t)^{2}\right\rangle
\simeq
\overline{|\alpha_{1}(t)-\alpha_{2}(t)|^{2}},
\end{equation}
where the overline denotes the average over stochastic trajectories.
No operator-ordering corrections are needed: the vacuum fluctuations are already accounted for by the random sampling of the initial conditions.
Concretely, each trajectory is initialized by drawing $\alpha_{k}(0)$ from the Wigner distribution of the initial product coherent state, i.e., $\alpha_{k}(0)=\alpha_{k,X}+\xi_{k}$ with $\xi_{k}$ a complex Gaussian random variable of zero mean and variance $\overline{|\xi_{k}|^{2}}=1/2$, and is then propagated according to Eq.~\eqref{Eq:TWA-SDE-N2}.
The synchronization measure is thus obtained as
\begin{equation}
\Ss_{c}^{\mathrm{TWA}}(t)=\frac{1}{\overline{|\alpha_{1}(t)-\alpha_{2}(t)|^{2}}},
\label{Eq:Sc-TWA}
\end{equation}
with the trajectory average evaluated from $10^{6}$ independent realizations integrated with the time step $\gamma_{\uparrow}dt=0.0025$.
Figure~\ref{Fig:N2TWA}(a) compares the resulting $\Ss_{c}^{\mathrm{TWA}}(t)$ with the exact master-equation dynamics for the two initial states $\rho_{A}$ and $\rho_{B}$ used in the main text.
The TWA reproduces the exact curves over the whole evolution and, in particular, captures the crossing of $\Ss_{c}^{A}(t)$ and $\Ss_{c}^{B}(t)$ that signals the synchronization Mpemba effect.
The deviation shown in Fig.~\ref{Fig:N2TWA}(b) remains at the level of $10^{-3}$ for both initial states, consistent with the statistical uncertainty of the finite trajectory ensemble.

The same construction gives the many-body synchronization measure directly from the TWA trajectories,
\begin{equation}
\Ss_{\perp}^{\mathrm{TWA}}(t)=
\left[
\frac{2}{N(N-1)}
\sum_{k<j}\overline{|\alpha_k(t)-\alpha_j(t)|^2}
\right]^{-1}.
\label{Eq:Sperp-TWA}
\end{equation}
This makes the TWA particularly convenient for evaluating the synchronization measure of large networks, where the exponential growth of the Hilbert space rules out an exact solution of the master equation.

\section{Experimental Proposal with a Continuous-Variable Quantum Circuit}

Continuous-variable (CV) quantum circuits provide a promising platform for realizing the SMpE: the required multimode Gaussian and non-Gaussian operations are within reach of current photonic and superconducting hardware, allowing the dissipative dynamics of coupled quantum vdP oscillators to be implemented directly at the circuit level.
To this end, we employ a quantum collision modell~\cite{Ciccarello2022,Karpat2019,Cattaneo2021a,Li2026c} that digitally reproduces the open dynamics: the evolution is decomposed into a sequence of short unitary interactions between the system oscillators and fresh ancillary modes, which recovers the master equation~\eqref{Eq:ME} in the continuous-time limit.
Below, we derive the collision-model representation and compile each collision step into an elementary CV gate sequence, so that the full dynamics is implemented by iterating a fixed circuit block.

\blueit{Quantum collision model.--} In the quantum collision model framework, the environment is represented by a stream of identical ancillary modes, each prepared in the same initial state $\eta$ and interacting with the system only once.
The dynamics over one time step $\delta t$ consists of the free evolution of the system followed by a single system-ancilla collision, described by the discrete-time map
\begin{equation}
    \rho_{S}^{n}\mapsto\rho_{S}^{n+1}=\mathrm{Tr}_{\eta_{n+1}}\left[U_{I}U_{S}\left(\rho_{S}^{n}\otimes\eta_{n+1}\right)U_{S}^{\dagger}U_{I}^{\dagger}\right],
    \label{Eq:collision-map}
\end{equation}
where $\rho_{S}^{n}$ is the system state after $n$ steps, $\eta_{n+1}$ denotes the joint state of the fresh ancilla modes entering the $(n+1)$th step, and the partial trace is taken over these ancillas after the collision.
The two unitaries play distinct roles: $U_{S}=e^{-iH\tau}$ is the free evolution of the system generated by the Hamiltonian $H=\omega\sum_{k}a_{k}^{\dagger}a_{k}$, while $U_{I}=e^{-iH_{I}\tau}$ describes the collision, with $H_{I}$ the interaction Hamiltonian between the system oscillators and the ancillas.
The master equation~\eqref{Eq:ME} contains three dissipative channels: the incoherent gain $\gamma_{\uparrow}\Ds[a_{k}^{\dagger}]$, the nonlinear loss $\gamma_{\downarrow}\Ds[a_{k}^{2}]$, and the pairwise dissipative coupling $\kappa_{kj}\Ds[a_{k}-a_{j}]$.
For concreteness, we present the collision-model construction for the minimal case of two coupled oscillators ($N=2$) with $\kappa_{12}\equiv\kappa$; the extension to larger networks is straightforward.
The gain and nonlinear-loss channels act independently on each oscillator, while the coupling channel acts on the pair.
Each collision step therefore involves five ancilla modes: $b_{k}$ and $c_{k}$ for the gain and nonlinear loss of oscillator $k=1,2$, and a single ancilla $d$ for the dissipative coupling.
The ancilla state entering every step factorizes as $\eta=\eta_{\gamma_{\uparrow}}\otimes\eta_{\gamma_{\downarrow}}\otimes\eta_{\kappa}$, with $\eta_{\gamma_{\uparrow}}=|0\rangle\langle0|_{b_{1}}\otimes|0\rangle\langle0|_{b_{2}}$, $\eta_{\gamma_{\downarrow}}=|0\rangle\langle0|_{c_{1}}\otimes|0\rangle\langle0|_{c_{2}}$, and $\eta_{\kappa}=|0\rangle\langle0|_{d}$, where the subscripts label the three channels.
Accordingly, the interaction Hamiltonian splits into three parts,
\begin{equation}
H_{I}=H_{\gamma_{\uparrow}}+H_{\gamma_{\downarrow}}+H_{\kappa},
\label{Eq:HI-total}
\end{equation}
where each term couples the system only to the ancilla modes of the corresponding channel.
The three interactions are given explicitly by
\begin{equation}
\begin{aligned}
H_{\gamma_{\uparrow}} &= \sqrt{\frac{\gamma_{\uparrow}}{\tau}}\sum_{k=1,2}\left(a_{k}^{\dagger}b_{k}^{\dagger}+a_{k}b_{k}\right),\\
H_{\gamma_{\downarrow}} &= \sqrt{\frac{\gamma_{\downarrow}}{\tau}}\sum_{k=1,2}\left(a_{k}^{2}c_{k}^{\dagger}+a_{k}^{\dagger 2}c_{k}\right),\\
H_{\kappa} &= \sqrt{\frac{\kappa}{\tau}}\left[\left(a_{1}-a_{2}\right)d^{\dagger}+\left(a_{1}^{\dagger}-a_{2}^{\dagger}\right)d\right].
\end{aligned}
\label{Eq:H-channels}
\end{equation}
Since the time step $\tau$ is small, we expand the unitaries $U_{S}$ and $U_{I}$, retaining all terms that contribute through first order in $\tau$, i.e.,
\begin{equation}
    \begin{aligned}
    U_{S}=&\mathrm{I}-iH\tau+O(\tau^{2}),\\
    U_{I}=&\mathrm{I}-iH_{I}\tau-\frac{1}{2}H_{I}^{2}\tau^{2}+O(\tau^{3/2}).
    \end{aligned}
\end{equation}
Accordingly, the joint unitary $U_{I}U_{S}$, retaining all contributions through first order in $\tau$, is given by
\begin{equation}
    U_{I}U_{S}=\mathrm{I}-i(H_{I}+H)\tau-\frac{1}{2}H_{I}^{2}\tau^{2}+O(\tau^{3/2}).
\end{equation}
The collision map~\eqref{Eq:collision-map} can then be evaluated by substituting this expansion and tracing over the ancilla modes.
For compactness, we denote the joint system-ancilla state before the collision by $X_n=\rho_S^n\otimes\eta$.
\begin{equation}
\begin{aligned}
\rho_{S}^{n+1}\simeq&\text{Tr}_{\eta}\left[(\mathrm{I}-i(H_{I}+H)\tau-\frac{1}{2}H_{I}^{2}\tau^{2})X_n(\mathrm{I}+i(H_{I}+H)\tau-\frac{1}{2}H_{I}^{2}\tau^{2})\right]+O(\tau^{3/2}),\\
=&\text{Tr}_{\eta}\left[X_n-i\tau(H_I+H)X_n+i\tau X_n(H_I+H)+\tau^2(H_I+H)X_n(H_I+H)-\frac{1}{2}\tau^2H_I^2X_n-\frac{1}{2}\tau^2X_nH_I^2\right]+O(\tau^{3/2}),\\
=&\text{Tr}_{\eta}\left[X_n-i\tau(H_I+H)X_n+i\tau X_n(H_I+H)+\tau^2H_I X_n H_I+\tau^2H_I X_n H+\tau^2H X_n H_I+\tau^2H X_n H-\frac{1}{2}\tau^2H_I^2X_n-\frac{1}{2}\tau^2X_nH_I^2\right]+O(\tau^{3/2}),\\
=&\text{Tr}_{\eta}\left[X_n-i\tau(H_I+H)X_n+i\tau X_n(H_I+H)+\tau^2H_I X_n H_I-\frac{1}{2}\tau^2H_I^2X_n-\frac{1}{2}\tau^2X_nH_I^2\right]+O(\tau^{3/2})\\
=&\rho_S^n-i\tau[H,\rho_S^n]-i\tau\text{Tr}_{\eta}\left(H_I X_n-X_nH_I\right)+\tau^2\text{Tr}_{\eta}\left[H_I X_n H_I-\frac{1}{2}H_I^2X_n-\frac{1}{2}X_nH_I^2\right]+O(\tau^{3/2})\\
=&\rho_S^n-i\tau[H,\rho_S^n]+\tau^2\text{Tr}_{\eta}\left[H_I X_n H_I-\frac{1}{2}H_I^2X_n-\frac{1}{2}X_nH_I^2\right]+O(\tau^{3/2}).
\end{aligned}
\label{Eq:map-expansion}
\end{equation}
Here, the terms $\tau^2H_I X_nH$, $\tau^2H X_nH_I$, and $\tau^2H X_nH$ are of orders $\tau^{3/2}$, $\tau^{3/2}$, and $\tau^2$, respectively, and are absorbed into the remainder only after the product has been fully expanded.
In the last equality, it is the partial trace of the term linear in $H_I$, rather than the joint operator itself, that vanishes:
\begin{equation}
\text{Tr}_{\eta}\left(H_I X_n-X_nH_I\right)=0.
\end{equation}
This follows because all ancillas are in the vacuum state and therefore satisfy $\text{Tr}_{\eta}(e\eta)=\text{Tr}_{\eta}(e^{\dagger}\eta)=0$ for $e\in\{b_1,b_2,c_1,c_2,d\}$.

We next evaluate the remaining terms that are quadratic in $H_I$.
Because $b_1$, $b_2$, $c_1$, $c_2$, and $d$ are independent vacuum modes, all cross terms between different interaction Hamiltonians vanish after the partial trace.  The same applies to terms involving different $b_k$ or different $c_k$ modes.
Therefore,
\begin{equation}
\begin{aligned}
\text{Tr}_{\eta}(H_I X_nH_I)
={}&\text{Tr}_{\eta}(H_{\gamma_{\uparrow}}X_nH_{\gamma_{\uparrow}})
+\text{Tr}_{\eta}(H_{\gamma_{\downarrow}}X_nH_{\gamma_{\downarrow}})
+\text{Tr}_{\eta}(H_{\kappa}X_nH_{\kappa}),\\
\text{Tr}_{\eta}(H_I^2X_n)
={}&\text{Tr}_{\eta}(H_{\gamma_{\uparrow}}^2X_n)
+\text{Tr}_{\eta}(H_{\gamma_{\downarrow}}^2X_n)
+\text{Tr}_{\eta}(H_{\kappa}^2X_n),\\
\text{Tr}_{\eta}(X_nH_I^2)
={}&\text{Tr}_{\eta}(X_nH_{\gamma_{\uparrow}}^2)
+\text{Tr}_{\eta}(X_nH_{\gamma_{\downarrow}}^2)
+\text{Tr}_{\eta}(X_nH_{\kappa}^2).
\end{aligned}
\end{equation}
Consequently, the three terms in the second-order contribution can be evaluated directly from Eq.~\eqref{Eq:H-channels}.
Using $\text{Tr}_{\eta}(b_kb_j^{\dagger}\eta)=\text{Tr}_{\eta}(c_kc_j^{\dagger}\eta)=\delta_{kj}$, $\text{Tr}_{\eta}(dd^{\dagger}\eta)=1$, and their normally ordered counterparts equal to zero, the gain channel becomes
\begin{equation}
\text{Tr}_{\eta}(H_{\gamma_{\uparrow}}X_nH_{\gamma_{\uparrow}})
 =\frac{\gamma_{\uparrow}}{\tau}\sum_{k=1,2}a_k^{\dagger}\rho_S^n a_k,\quad
\text{Tr}_{\eta}(H_{\gamma_{\uparrow}}^2X_n)
 =\frac{\gamma_{\uparrow}}{\tau}\sum_{k=1,2}a_k a_k^{\dagger}\rho_S^n,\quad
\text{Tr}_{\eta}(X_nH_{\gamma_{\uparrow}}^2)
 =\frac{\gamma_{\uparrow}}{\tau}\sum_{k=1,2}\rho_S^n a_k a_k^{\dagger}.
\end{equation}
Similarly, the nonlinear-loss channel gives
\begin{equation}
\text{Tr}_{\eta}(H_{\gamma_{\downarrow}}X_nH_{\gamma_{\downarrow}})
 =\frac{\gamma_{\downarrow}}{\tau}\sum_{k=1,2}a_k^2\rho_S^n a_k^{\dagger 2},\quad
\text{Tr}_{\eta}(H_{\gamma_{\downarrow}}^2X_n)
 =\frac{\gamma_{\downarrow}}{\tau}\sum_{k=1,2}a_k^{\dagger 2}a_k^2\rho_S^n,\quad
\text{Tr}_{\eta}(X_nH_{\gamma_{\downarrow}}^2)
 =\frac{\gamma_{\downarrow}}{\tau}\sum_{k=1,2}\rho_S^n a_k^{\dagger 2}a_k^2.
\end{equation}
For the dissipative-coupling channel, we find
\begin{equation}
\text{Tr}_{\eta}(H_{\kappa}X_nH_{\kappa})
 =\frac{\kappa}{\tau}(a_1-a_2)\rho_S^n(a_1^{\dagger}-a_2^{\dagger}),\quad
\text{Tr}_{\eta}(H_{\kappa}^2X_n)
 =\frac{\kappa}{\tau}(a_1^{\dagger}-a_2^{\dagger})(a_1-a_2)\rho_S^n,\quad
\text{Tr}_{\eta}(X_nH_{\kappa}^2)
 =\frac{\kappa}{\tau}\rho_S^n(a_1^{\dagger}-a_2^{\dagger})(a_1-a_2).
\end{equation}
Substituting these results into the second-order contribution in Eq.~\eqref{Eq:map-expansion} gives
\begin{equation}
\tau^2\text{Tr}_{\eta}\left[H_I X_n H_I-\frac{1}{2}H_I^2X_n-\frac{1}{2}X_nH_I^2\right]
={}\tau\left\{\sum_{k=1,2}\left(\gamma_{\uparrow}\Ds[a_k^{\dagger}]\rho_S^n+\gamma_{\downarrow}\Ds[a_k^2]\rho_S^n\right)+\kappa\Ds[a_1-a_2]\rho_S^n\right\}.
\end{equation}
Substituting this result into Eq.~\eqref{Eq:map-expansion}, we obtain the discrete reduced dynamics
\begin{equation}
\begin{aligned}
\rho_S^{n+1}=\rho_S^n+\tau\left\{-i[H,\rho_S^n]+\sum_{k=1,2}\left(\gamma_{\uparrow}\Ds[a_k^{\dagger}]\rho_S^n+\gamma_{\downarrow}\Ds[a_k^2]\rho_S^n\right)+\kappa\Ds[a_1-a_2]\rho_S^n\right\}+O(\tau^{3/2}).
\end{aligned}
\end{equation}
Finally, subtracting $\rho_S^n$ from both sides, dividing by $\tau$, and taking the limit $\tau\rightarrow0$ yields
\begin{equation}
\frac{d\rho_S}{dt}=-i[H,\rho_S]+\sum_{k=1,2}\left(\gamma_{\uparrow}\Ds[a_k^{\dagger}]\rho_S+\gamma_{\downarrow}\Ds[a_k^2]\rho_S\right)+\kappa\Ds[a_1-a_2]\rho_S,
\end{equation}
which is precisely Eq.~\eqref{Eq:ME} for two coupled vdP oscillators.

\blueit{Continuous-variable circuit decomposition.--} Having established the continuous-time limit, we approximate the interaction unitary at finite $\tau$ by the channel-resolved sequence
\begin{equation}
U_I=e^{-i(H_{\gamma_{\uparrow}}+H_{\gamma_{\downarrow}}+H_{\kappa})\tau}
\simeq U_I^{\mathrm{circ}}\equiv U_{\kappa}U_{\gamma_{\downarrow}}U_{\gamma_{\uparrow}},
\qquad
U_{\mathrm{step}}=U_I^{\mathrm{circ}}U_S.
\end{equation}
The approximation is not an equality at finite $\tau$; it becomes equivalent to $U_I$ at the level of the reduced generator as $\tau\to0$.

Using the PennyLane gate conventions~\cite{Bergholm2022}, the free rotation is implemented as
\begin{equation}
U_S\ \longleftrightarrow
\prod_{k=1,2}\texttt{qml.Rotation}(-\omega\tau,\,\texttt{wires}=a_k),
\end{equation}
because \texttt{qml.Rotation}$(\phi)=e^{i\phi a^{\dagger}a}$.
The parametric-amplification operation is implemented by
\begin{equation}
U_{\gamma_{\uparrow}}\ \longleftrightarrow\
\prod_{k=1,2}\texttt{qml.TwoModeSqueezing}
\left(\sqrt{\gamma_{\uparrow}\tau},\frac{\pi}{2},\texttt{wires}=[a_k,b_k]\right),
\end{equation}
where the phase $\pi/2$ follows from
$\texttt{qml.TwoModeSqueezing}(r,\pi/2)=e^{-ir(a^{\dagger}b^{\dagger}+ab)}$.

Finally, $U_{\kappa}$ is implemented by the PennyLane sequence
\begin{equation}
\begin{aligned}
&\texttt{qml.Beamsplitter}\left(\frac{\pi}{4},0,\texttt{wires}=[a_1,a_2]\right),\\
&\texttt{qml.Beamsplitter}\left(-\sqrt{2\kappa\tau},\frac{\pi}{2},\texttt{wires}=[a_1,d]\right),\\
&\texttt{qml.Beamsplitter}\left(-\frac{\pi}{4},0,\texttt{wires}=[a_1,a_2]\right).
\end{aligned}
\end{equation}
The first and last gates transform to and from the relative-mode basis, while the middle gate realizes the required coupling to $d$.

The nonlinear-loss operation is the most demanding part of the circuit.  It is not a
native PennyLane gate: each factor is a two-mode cubic interaction,
\begin{equation}
U_{\gamma_{\downarrow}}
=\prod_{k=1,2}e^{-i\sqrt{\gamma_{\downarrow}\tau}(a_k^2c_k^{\dagger}+a_k^{\dagger 2}c_k)}.
\end{equation}
It is not identical to a single \texttt{qml.CubicPhase} operation, but it can be
compiled from Gaussian gates and cubic-phase gates.  Using the quadrature
convention introduced above and the analogous ancilla quadratures
$q_{c_k},p_{c_k}$, its generator is
\begin{equation}
a_k^2c_k^{\dagger}+a_k^{\dagger 2}c_k
=\frac{1}{\sqrt{2}}\left[(q_k^2-p_k^2)q_{c_k}+(q_kp_k+p_kq_k)p_{c_k}\right].
\end{equation}
For each pair $(a_k,c_k)$, the required cubic polynomial is decomposed into cubes
of linear quadratures as
\begin{align}
q_k^2q_{c_k}&=\frac{(2q_k+q_{c_k})^3-(2q_k-q_{c_k})^3-2q_{c_k}^3}{24},\\
p_k^2q_{c_k}&=\frac{(2p_k+q_{c_k})^3-(2p_k-q_{c_k})^3-2q_{c_k}^3}{24},\\
(q_kp_k+p_kq_k)p_{c_k}
&=\frac{(q_k+p_k+p_{c_k})^3-(q_k+p_k-p_{c_k})^3}{12}-\frac{(q_k-p_k+p_{c_k})^3+(-q_k+p_k+p_{c_k})^3}{12}.
\end{align}
Each linear quadrature $Q_j$ in this decomposition can be mapped to the position
quadrature of one mode by a Gaussian circuit $G_j$, assembled from operations such
as \texttt{qml.Rotation}, \texttt{qml.Squeezing}, and
\texttt{qml.Beamsplitter}.  Consequently,
\begin{equation}
G_j^{\dagger}\,\texttt{qml.CubicPhase}(\gamma_j)\,G_j
=\exp\left(i\frac{\gamma_j}{3}Q_j^3\right),
\end{equation}
where the gate acts on the mode selected by $G_j$.  If the above decomposition is
written as $a_k^2c_k^{\dagger}+a_k^{\dagger2}c_k=\sum_jg_jQ_j^3$, a first-order
Trotter decomposition of each nonlinear factor is
\begin{equation}
e^{-i\sqrt{\gamma_{\downarrow}\tau}(a_k^2c_k^{\dagger}+a_k^{\dagger2}c_k)}
\simeq
\left[\prod_jG_j^{\dagger}\,\texttt{qml.CubicPhase}\left(-\frac{3\sqrt{\gamma_{\downarrow}\tau}\,g_j}{M}\right)G_j\right]^M,
\qquad M\gg1,
\end{equation}
where the coefficients $g_j$ are read directly from the preceding identities.  Thus, the nonlinear channel can be implemented by a Gaussian network interleaved with several cubic-phase gates.
After each step, the five ancillas are discarded and replaced by fresh vacuum modes; repeating the circuit $n$ times gives $t=n\tau$.

\section{Analytical crossing time for the synchronization Mpemba effect}

The crossing time $t_{\times}$ is the central observable of the SMpE, and its dependence on the coupling and damping rates encodes the dynamical origin of the effect.
Here we develop an analytical estimate of $t_{\times}$ within a unified modal framework valid for networks of any size $N$, and then specialize it to the two-oscillator case, where the synchronization dynamics is carried by a single relative mode, and to genuine networks, where the transverse modes relax independently at rates set by the graph-Laplacian eigenvalues.
The nonlinear two-photon loss places the exact dynamics beyond analytical reach; the key observation that makes an analytical treatment possible is that the transverse modes remain nearly Gaussian at all times, which reduces the synchronization dynamics to a tractable linear problem.

\blueit{Laplacian eigenbasis.--}
We start from the mean-field equations~\eqref{Eq:MF-Laplacian}.
For an undirected network, the graph Laplacian is real symmetric and admits the spectral decomposition
\begin{equation}
L=\sum_{\mu=1}^{N}\lambda_\mu\,\boldsymbol{u}_\mu\boldsymbol{u}_\mu^{\mathrm T},
\qquad
\boldsymbol{u}_\mu^{\mathrm T}\boldsymbol{u}_\nu=\delta_{\mu\nu},
\qquad
0=\lambda_1\leq\lambda_2\leq\cdots\leq\lambda_N .
\label{Eq:Unified-L-spectral}
\end{equation}
Since each row of $L$ sums to zero, the uniform vector $\boldsymbol{u}_1=N^{-1/2}(1,\ldots,1)^{\mathrm T}$ is a zero mode, $L\boldsymbol{u}_1=0$, which represents the collective synchronized mode.
Moreover, the quadratic form $\boldsymbol{v}^{\mathrm T}L\boldsymbol{v}=\sum_{k<j}\kappa_{kj}(v_k-v_j)^2\geq0$ [cf.~Eq.~\eqref{Eq:L-quadratic}] shows that all eigenvalues are nonnegative, and for a connected network the zero eigenvalue is simple, $\lambda_2>0$.
The remaining eigenvectors $\boldsymbol{u}_{\mu}$, $\mu=2,\ldots,N$, form an orthonormal basis of the transverse subspace, which is directly damped by the nonlocal dissipation.

We expand the vector of complex amplitudes in this eigenbasis [cf.~Eq.~\eqref{Eq:mode-quadratures}] and remove the bare rotation and the initial collective phase by introducing the co-rotating modal amplitudes
\begin{equation}
\boldsymbol{\alpha}(t)=\sum_{\mu=1}^{N}\widetilde{\alpha}_\mu(t)\boldsymbol{u}_\mu,
\qquad
\widetilde{\alpha}_\mu=\boldsymbol{u}_\mu^{\mathrm T}\boldsymbol{\alpha},
\qquad
\beta_\mu(t)=e^{i\omega t-i\phi_0}\widetilde{\alpha}_\mu(t),
\label{Eq:Unified-beta}
\end{equation}
where $\phi_0=\arg\widetilde{\alpha}_1(0)$ is the initial phase of the collective mode, fixed below.
Projecting the vector mean-field equation~\eqref{Eq:MF-Laplacian-vector} onto $\boldsymbol{u}_\mu$, using $\boldsymbol{u}_\mu^{\mathrm T}L=\lambda_\mu\boldsymbol{u}_\mu^{\mathrm T}$, gives
\begin{equation}
\dot{\widetilde{\alpha}}_\mu
=\left(-i\omega+\frac{\gamma_{\uparrow}}{2}-\frac{\lambda_\mu}{2}\right)\widetilde{\alpha}_\mu
-\gamma_{\downarrow}\sum_{k=1}^{N}u_{\mu k}|\alpha_k|^2\alpha_k .
\label{Eq:Unified-projected-lab}
\end{equation}
The inverse transformation and the co-rotating definition give
\begin{equation}
\alpha_k(t)=e^{-i\omega t+i\phi_0}\sum_{\nu=1}^{N}u_{k\nu}\beta_\nu(t),
\qquad
\dot\beta_\mu=e^{i\omega t-i\phi_0}\bigl(\dot{\widetilde{\alpha}}_\mu+i\omega\widetilde{\alpha}_\mu\bigr).
\label{Eq:Unified-inverse}
\end{equation}
The common phase cancels from the cubic term in Eq.~\eqref{Eq:Unified-projected-lab}, so the mean-field equation in the co-rotating basis reads
\begin{equation}
\dot{\beta}_\mu
=\frac{\gamma_{\uparrow}-\lambda_\mu}{2}\,\beta_\mu
-\gamma_{\downarrow}\sum_{k=1}^{N}u_{\mu k}
\left|\sum_{\nu=1}^{N}u_{k\nu}\beta_\nu\right|^2
\left(\sum_{\sigma=1}^{N}u_{k\sigma}\beta_\sigma\right).
\label{Eq:Unified-projected}
\end{equation}
The Laplacian coupling is diagonal in this basis, with the collective mode unaffected by the nonlocal dissipation and each transverse mode damped at the rate $\lambda_\mu/2$; the local nonlinear loss, however, mixes different Laplacian modes away from the synchronized manifold.

\begin{figure}[!htpb]
    \centering
    \includegraphics[width=0.9\textwidth]{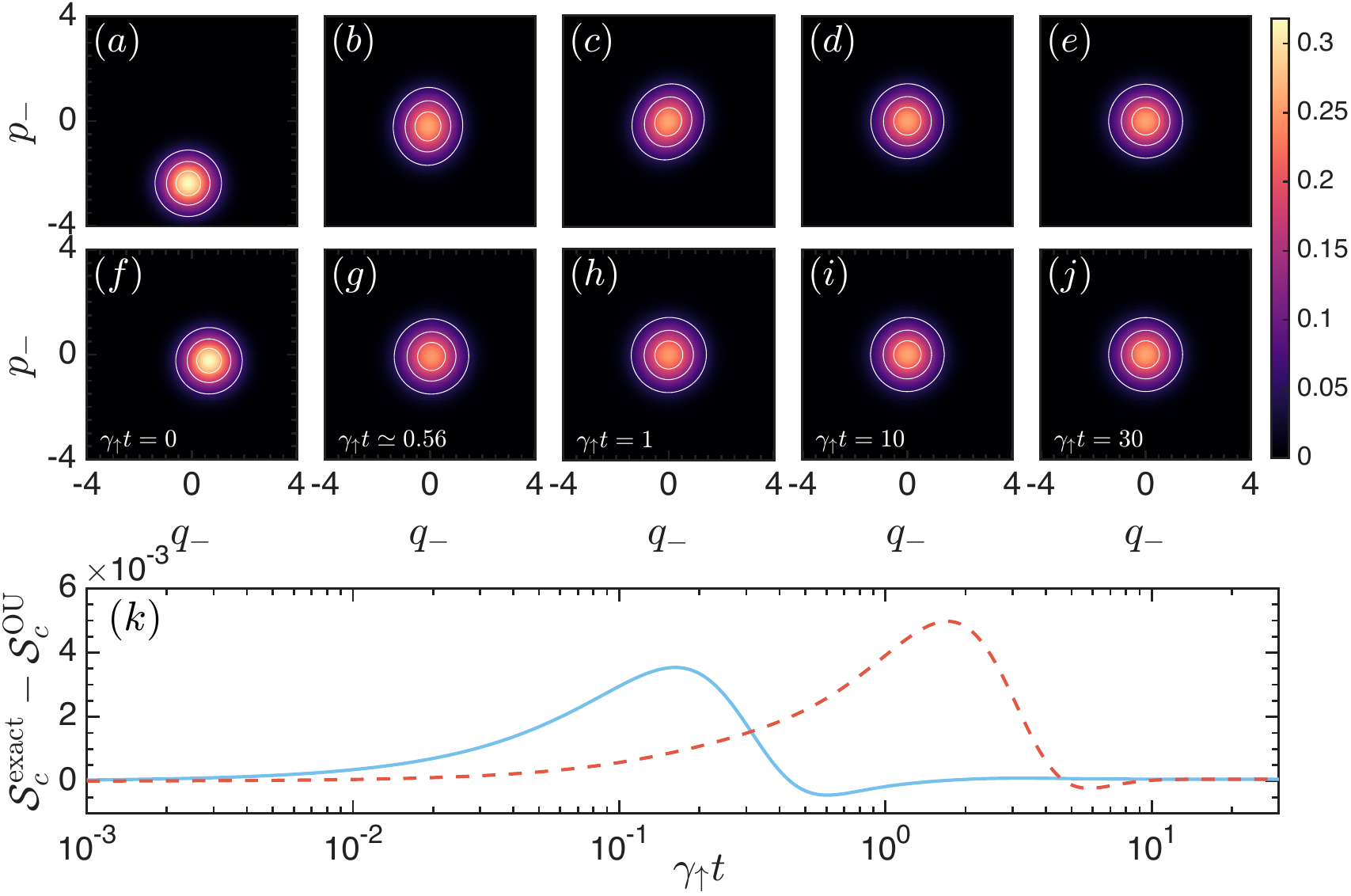}
    \caption{\label{Fig:N2RelatMW} Relative-mode Wigner functions and the Gaussian reference dynamics.
    (a)--(e) The relative-mode Wigner function $W(q_{-},p_{-})$ for the initial state $\rho_{A}$ at $\gamma_{\uparrow}t=0$, $0.56$, $1$, $10$, and $30$.
    (f)--(j) The same for the initial state $\rho_{B}$. The two initial states are identical to those used in the main text. In each panel, the white solid lines are contour lines of the reference Gaussian state that shares the same first and second moments as $\rho(t)$. (k) Deviation of the exact synchronization measure from the OU result, $\Ss_{c}^{\mathrm{exact}}-\Ss_{c}^{\mathrm{OU}}$, for $\rho_A$ (blue solid) and $\rho_B$ (red dashed). The parameters are $\omega=\gamma_{\uparrow}$, $\gamma_{\downarrow}=0.1\gamma_{\uparrow}$, and $\kappa=4\gamma_{\uparrow}$.}
\end{figure}

\blueit{Gaussian closure.--}
Equation~\eqref{Eq:Unified-projected} involves no approximation beyond the mean-field factorization; nevertheless, because of the nonlinear mode mixing, it is not analytically tractable in general.
The key observation is that the transverse modes remain nearly Gaussian at all times.
This near-Gaussianity has a clear physical origin: each transverse mode describes an error direction orthogonal to the synchronization manifold and is the combination directly damped by the nonlocal dissipation, at the rate $\lambda_\mu/2$.
In the synchronizing regime, this restoring dissipation confines the transverse fluctuations to the vicinity of the synchronized manifold, so that the nonlinearity enters only through small corrections beyond linear order in the transverse amplitudes and through the coefficients of the linearized dynamics, which are set by the collective mode as it relaxes toward the limit cycle.
Moreover, initial product coherent states remain product coherent states in the Laplacian modal basis, so every transverse mode starts exactly Gaussian, and linear Markovian dynamics preserves this Gaussianity.
We therefore linearize the transverse dynamics about the synchronized background and restore the quantum noise injected by the dissipative channels at the same semiclassical order, which promotes each transverse mode to a time-dependent Ornstein--Uhlenbeck (OU) process.
The accuracy of this closure is verified against the exact master equation for the two-oscillator case below [Fig.~\ref{Fig:N2RelatMW}].

\blueit{Collective mode.--}
For a connected network, complete synchronization, $\alpha_1=\cdots=\alpha_N$, means that the amplitude vector is proportional to $\boldsymbol{u}_1$, i.e., $\widetilde{\alpha}_{\mu\geq2}=0$.
In the co-rotating frame the collective amplitude can then be chosen real and nonnegative, $\beta_1=\sqrt{n_1}$, with $n_1=|\widetilde{\alpha}_1|^2$.
On the synchronized manifold, $\alpha_k=\widetilde{\alpha}_1/\sqrt N$, and hence $\sum_k u_{1k}|\alpha_k|^2\alpha_k=|\widetilde{\alpha}_1|^2\widetilde{\alpha}_1/N$; projecting Eq.~\eqref{Eq:Unified-projected-lab} onto the uniform vector gives
\begin{equation}
\dot{\widetilde{\alpha}}_1
=\left(-i\omega+\frac{\gamma_{\uparrow}}{2}\right)\widetilde{\alpha}_1
-\frac{\gamma_{\downarrow}}{N}|\widetilde{\alpha}_1|^2\widetilde{\alpha}_1 .
\label{Eq:Unified-collective-amplitude}
\end{equation}
Thus, to leading order in the transverse amplitudes, the collective mode evolves independently as a single vdP oscillator with a rescaled nonlinear loss.
Writing $\widetilde{\alpha}_1=\sqrt{n_1}e^{i\phi_1}$ in Eq.~\eqref{Eq:Unified-collective-amplitude} yields
\begin{equation}
\dot n_1=\gamma_{\uparrow}n_1-\frac{2\gamma_{\downarrow}}{N}n_1^2,
\qquad
\dot\phi_1=-\omega .
\label{Eq:Unified-collective-polar}
\end{equation}
Choosing $\phi_0=\phi_1(0)$, the collective mode is therefore $\widetilde{\alpha}_1(t)=\sqrt{n_1(t)}e^{-i\omega t+i\phi_0}$ and $\beta_1(t)=\sqrt{n_1(t)}$.
The radial equation is logistic [cf.~Eq.~\eqref{Eq:MF-logistic}], with the exact solution
\begin{equation}
n_1(t)=\frac{n_{1,\text{ss}}}{1+c e^{-\gamma_{\uparrow}t}},
\qquad
n_{1,\text{ss}}=\frac{N\gamma_{\uparrow}}{2\gamma_{\downarrow}}=Nr_{\text{LC}}^2,
\qquad
c=\frac{n_{1,\text{ss}}}{n_1(0)}-1,
\label{Eq:Unified-logistic}
\end{equation}
where $n_{1,\text{ss}}$ is the steady intensity of the collective mode on the synchronized limit cycle [cf.~Eq.~\eqref{Eq:rlc}].
For later use, we introduce the rescaled collective intensity
\begin{equation}
z(t)\equiv\frac{n_1(t)}{n_{1,\text{ss}}}
=\frac{e^{\gamma_{\uparrow}t}}{e^{\gamma_{\uparrow}t}+c},
\qquad
\dot z=\gamma_{\uparrow}z(1-z),
\label{Eq:Unified-z}
\end{equation}
which starts from $z(0)=n_1(0)/n_{1,\text{ss}}=1/(1+c)$ and approaches $1$ on the gain timescale $\gamma_{\uparrow}^{-1}$ as the collective mode approaches the limit cycle; correspondingly, $c>0$ ($c<0$) when the collective mode starts below (above) the limit cycle.

\blueit{Linearized transverse dynamics.--}
We now expand the nonlinear term of Eq.~\eqref{Eq:Unified-projected} about this time-dependent synchronized background to first order in the transverse amplitudes.
Writing $\sum_{\nu}u_{k\nu}\beta_\nu=u_{k1}\sqrt{n_1}+\sum_{\nu\geq2}u_{k\nu}\beta_\nu$ and using $u_{k1}=1/\sqrt N$ gives
\begin{equation}
\left|\sum_{\nu=1}^{N}u_{k\nu}\beta_\nu\right|^2
\left(\sum_{\sigma=1}^{N}u_{k\sigma}\beta_\sigma\right)
=\frac{u_{k1}}{N}\,n_1^{3/2}
+\frac{n_1}{N}\sum_{\nu=2}^{N}u_{k\nu}\bigl(2\beta_\nu+\beta_\nu^*\bigr)
+\Os(\beta^{2}).
\label{Eq:Unified-nonlinear-expansion}
\end{equation}
Upon projection onto $\boldsymbol{u}_\mu$ with $\mu\geq2$, the first term vanishes by orthogonality, $\sum_k u_{\mu k}u_{k1}=0$, while the second reduces to $(n_1/N)(2\beta_\mu+\beta_\mu^*)$.
The linearized transverse-mode equations thus read
\begin{equation}
\dot\beta_\mu
=-\left[\frac{\lambda_\mu}{2}-\frac{\gamma_{\uparrow}}{2}+\frac{2\gamma_{\downarrow}}{N}n_1(t)\right]\beta_\mu
-\frac{\gamma_{\downarrow}}{N}n_1(t)\beta_\mu^*,
\qquad \mu\geq2 .
\label{Eq:Unified-linear-beta}
\end{equation}
Since the coefficients are real, the co-rotating quadratures of each transverse mode, $x_\mu=(\beta_\mu+\beta_\mu^*)/\sqrt2$ and $y_\mu=-i(\beta_\mu-\beta_\mu^*)/\sqrt2$, decouple,
\begin{equation}
\dot x_\mu=-\left[\frac{\lambda_\mu}{2}-\frac{\gamma_{\uparrow}}{2}+\frac{3\gamma_{\downarrow}}{N}n_1(t)\right]x_\mu,
\qquad
\dot y_\mu=-\left[\frac{\lambda_\mu}{2}-\frac{\gamma_{\uparrow}}{2}+\frac{\gamma_{\downarrow}}{N}n_1(t)\right]y_\mu,
\qquad \mu\geq2,
\label{Eq:Unified-xy}
\end{equation}
with $x_\mu$ the error along the limit-cycle phase, i.e., the amplitude mismatch, and $y_\mu$ the perpendicular error, i.e., the phase difference.

\blueit{Langevin dynamics and second moments.--}
The drift equations~\eqref{Eq:Unified-xy} capture only the deterministic part of the Lindblad generator: each dissipative channel also injects Markovian quantum noise, without which the transverse fluctuations would decay to zero rather than to the finite synchronization error imposed by the vacuum fluctuations.
At the same leading semiclassical order as the mean-field background, i.e., a Gaussian closure in which operator-ordering corrections are discarded, the noise strengths are evaluated on the synchronized background $|\alpha_k|^2=n_1(t)/N$.
In the Laplacian basis the nonlocal channel is diagonal, and on the synchronized background the local noise is the same at every site; consequently, different transverse modes do not mix at this order, and each obeys an independent Langevin equation,
\begin{equation}
dx_\mu=-a_{x,\mu}(t)x_\mu\,dt+\sqrt{D_\mu(t)}\,dW_{x,\mu},
\qquad
dy_\mu=-a_{y,\mu}(t)y_\mu\,dt+\sqrt{D_\mu(t)}\,dW_{y,\mu},
\label{Eq:Unified-OU}
\end{equation}
with independent real Wiener increments $\langle dW_{r,\mu}dW_{s,\nu}\rangle=\delta_{rs}\delta_{\mu\nu}\,dt$, drift rates read off from Eq.~\eqref{Eq:Unified-xy},
\begin{equation}
a_{x,\mu}(t)=\frac{\lambda_\mu-\gamma_{\uparrow}}{2}+\frac{3\gamma_{\downarrow}}{N}n_1(t),
\qquad
a_{y,\mu}(t)=\frac{\lambda_\mu-\gamma_{\uparrow}}{2}+\frac{\gamma_{\downarrow}}{N}n_1(t),
\label{Eq:Unified-drift-rates}
\end{equation}
and the diffusion rate
\begin{equation}
D_\mu(t)=\frac{\lambda_\mu}{2}+\frac{\gamma_{\uparrow}}{2}+\frac{2\gamma_{\downarrow}}{N}n_1(t),
\label{Eq:Unified-Dmu}
\end{equation}
where the three terms originate, respectively, from the nonlocal coupling, the local gain, and the two-photon loss evaluated on the collective background.
Each transverse mode thus follows a time-dependent OU process; Gaussianity is a property of this linearized approximation, not of the full nonlinear master equation.

For this OU process, the synchronization error is entirely determined by the second moments $X_\mu=\langle x_\mu^2\rangle$ and $Y_\mu=\langle y_\mu^2\rangle$: the co-rotation acts as an orthogonal transformation on each quadrature pair $(Q_\mu,P_\mu)$ of Eq.~\eqref{Eq:mode-quadratures} and therefore leaves $Q_\mu^2+P_\mu^2$ invariant, so that $\langle Q_\mu^2+P_\mu^2\rangle=X_\mu+Y_\mu$.
Applying It\^o's rule to Eq.~\eqref{Eq:Unified-OU} and taking the ensemble average gives $\dot X_\mu=-2a_{x,\mu}X_\mu+D_\mu$ and $\dot Y_\mu=-2a_{y,\mu}Y_\mu+D_\mu$; substituting $2\gamma_{\downarrow}n_1/N=\gamma_{\uparrow}z(t)$ and introducing
\begin{equation}
\lambda_{0,\mu}\equiv\lambda_\mu-\gamma_{\uparrow},
\qquad
q_{0,\mu}\equiv\frac{\lambda_\mu+\gamma_{\uparrow}}{2},
\label{Eq:Unified-rates}
\end{equation}
both equations take the common form
\begin{equation}
\dot M_{\mu,n}=-\bigl[\lambda_{0,\mu}+n\gamma_{\uparrow}z(t)\bigr]M_{\mu,n}+q_{0,\mu}+\gamma_{\uparrow}z(t),
\qquad n=1,3,
\label{Eq:Unified-Mn}
\end{equation}
with $M_{\mu,3}=X_\mu$ and $M_{\mu,1}=Y_\mu$.
Thus, the only difference between the amplitude and phase errors is the factor of three versus one multiplying the collective intensity in the drift.

\blueit{Exact solution.--}
Equation~\eqref{Eq:Unified-Mn} is linear with time-dependent coefficients and admits an exact solution.
The required integrating factors follow directly from the logistic solution~\eqref{Eq:Unified-z}: since $\int_0^t\gamma_{\uparrow}z(s)\,ds=\ln[(e^{\gamma_{\uparrow}t}+c)/(1+c)]$, we define
\begin{equation}
R(t)\equiv\frac{1+c}{e^{\gamma_{\uparrow}t}+c},
\qquad R(0)=1,
\label{Eq:Unified-R}
\end{equation}
in terms of which the homogeneous solution of Eq.~\eqref{Eq:Unified-Mn} reads
\begin{equation}
M_{\mu,n}^{\mathrm{(h)}}(t)=M_{\mu,n}(0)\,e^{-\lambda_{0,\mu}t}R^n(t).
\label{Eq:Unified-homogeneous}
\end{equation}
A particular solution follows from the observation that the Bernstein basis functions $\varphi_{n,k}(z)=z^k(1-z)^{n-k}$ obey
\begin{equation}
\frac{d}{dt}\varphi_{n,k}[z(t)]=\gamma_{\uparrow}\bigl(k-nz(t)\bigr)\varphi_{n,k}[z(t)],
\label{Eq:Unified-Bernstein-derivative}
\end{equation}
as follows from $\dot z=\gamma_{\uparrow}z(1-z)$.
Substituting the ansatz $P_{n,\mu}[z]=\sum_{k=0}^{n}\binom{n}{k}A_{n,k,\mu}\varphi_{n,k}(z)$ into Eq.~\eqref{Eq:Unified-Mn} and using the partition of unity $\sum_k\binom{n}{k}\varphi_{n,k}(z)=1$ and $\sum_k k\binom{n}{k}\varphi_{n,k}(z)=nz$ gives
\begin{equation}
\left(\frac{d}{dt}+\lambda_{0,\mu}+n\gamma_{\uparrow}z\right)P_{n,\mu}[z(t)]
=\sum_{k=0}^{n}\binom{n}{k}\bigl(\lambda_{0,\mu}+k\gamma_{\uparrow}\bigr)A_{n,k,\mu}\,\varphi_{n,k}(z)
=q_{0,\mu}+\gamma_{\uparrow}z(t),
\end{equation}
which fixes $A_{n,k,\mu}=(q_{0,\mu}+k\gamma_{\uparrow}/n)/(\lambda_{0,\mu}+k\gamma_{\uparrow})$ and hence
\begin{equation}
P_{n,\mu}[z]
=\sum_{k=0}^{n}\binom{n}{k}\,
\frac{q_{0,\mu}+k\gamma_{\uparrow}/n}{\lambda_{0,\mu}+k\gamma_{\uparrow}}\,
z^k(1-z)^{n-k},
\qquad n=1,3.
\label{Eq:Unified-Pn}
\end{equation}
Variation of constants then yields the full exact solution
\begin{equation}
M_{\mu,n}(t)=P_{n,\mu}[z(t)]+\bigl[M_{\mu,n}(0)-P_{n,\mu}[z(0)]\bigr]e^{-\lambda_{0,\mu}t}R^n(t),
\qquad n=1,3,
\label{Eq:Unified-Mn-solution}
\end{equation}
with the constant multiplying the homogeneous solution fixed by the initial condition.
Explicitly, the second moments of transverse mode $\mu$ are
\begin{equation}
\begin{aligned}
X_\mu(t)&=P_{3,\mu}[z(t)]+\bigl[X_\mu(0)-P_{3,\mu}[z(0)]\bigr]e^{-\lambda_{0,\mu}t}R^3(t),\\
Y_\mu(t)&=P_{1,\mu}[z(t)]+\bigl[Y_\mu(0)-P_{1,\mu}[z(0)]\bigr]e^{-\lambda_{0,\mu}t}R(t).
\end{aligned}
\label{Eq:Unified-XY-solution}
\end{equation}
For product coherent initial states with amplitudes $\alpha_k(0)$, each transverse mode starts in a displaced vacuum state, so
\begin{equation}
X_\mu(0)=\frac12+2\bigl[\mathrm{Re}\,\beta_\mu(0)\bigr]^2,
\qquad
Y_\mu(0)=\frac12+2\bigl[\mathrm{Im}\,\beta_\mu(0)\bigr]^2,
\qquad
\beta_\mu(0)=e^{-i\phi_0}\boldsymbol{u}_\mu^{\mathrm T}\boldsymbol{\alpha}(0),
\label{Eq:Unified-initial-moments}
\end{equation}
with $1/2$ the vacuum variance.
Equations~\eqref{Eq:Unified-Pn}--\eqref{Eq:Unified-initial-moments} are the common analytical input for both the two-oscillator and many-body discussions below; for two preparations $j=A,B$ with different collective backgrounds, the same formulas apply with $c$, $z$, and $R$ replaced by $c_j$, $z_j$, and $R_j$.

\blueit{General crossing condition.--}
The crossing time of two preparations $A$ and $B$ is fixed by the equality of their synchronization errors.
Within the OU description, the error decomposes over the transverse modes, and we define for each mode the background difference $\Bs_\mu(t)$ and the transient part $\Hs_\mu(t)$ by
\begin{equation}
\begin{aligned}
\Bs_\mu(t)&=\sum_{n=1,3}\bigl(P_{n,\mu}[z_A(t)]-P_{n,\mu}[z_B(t)]\bigr),\\
\Hs_\mu(t)&=\sum_{n=1,3}\Bigl(\bigl[M^A_{\mu,n}(0)-P_{n,\mu}[z_A(0)]\bigr]R_A^n(t)-\bigl[M^B_{\mu,n}(0)-P_{n,\mu}[z_B(0)]\bigr]R_B^n(t)\Bigr),
\end{aligned}
\label{Eq:Unified-BH}
\end{equation}
in terms of which the error difference between the two states reads
\begin{equation}
\Delta D_\perp(t)=\frac{1}{N-1}\sum_{\mu=2}^{N}\bigl[\Bs_\mu(t)+e^{-\lambda_{0,\mu}t}\Hs_\mu(t)\bigr],
\qquad
\Delta D_\perp(t_\times)=0 .
\label{Eq:Unified-F}
\end{equation}
Since both states relax to the same steady state, $\Delta D_\perp(t)\to0$ as $t\to\infty$; taking $\rho_A$ to be the initially less synchronized state, $\Delta D_\perp(0)>0$, an approach from below then guarantees at least one crossing, in fact an odd number of them, so the existence of the SMpE is decided by the sign of $\Delta D_\perp(t)$ at late times.
At late times, when the transients $e^{-\lambda_{0,\mu}t}\Hs_\mu(t)$ have decayed, expanding $z_j(t)=1-c_je^{-\gamma_{\uparrow}t}+\Os(e^{-2\gamma_{\uparrow}t})$ reduces each background difference to
\begin{equation}
\Bs_\mu(t)\simeq Q_\mu'[1]\,(c_B-c_A)\,e^{-\gamma_{\uparrow}t},
\qquad
Q_\mu[z]\equiv P_{3,\mu}[z]+P_{1,\mu}[z].
\label{Eq:Unified-B-late}
\end{equation}
The approach is thus from below whenever $(c_B-c_A)\sum_{\mu=2}^{N}Q_\mu'[1]<0$, which, together with $\Delta D_\perp(0)>0$, constitutes a screening criterion for the SMpE: it involves only the system rates, the Laplacian spectrum, and the initial coherent amplitudes, so candidate states can be screened without solving the master equation and verified afterward against the exact dynamics.

\blueit{The $N=2$ case.--}
For two oscillators with coupling strength $\kappa\equiv\kappa_{12}$, $L=\kappa\bigl(\begin{smallmatrix}1&-1\\-1&1\end{smallmatrix}\bigr)$, so that $\lambda_2=2\kappa$ and the only transverse mode is the relative mode $\beta_2\equiv\beta_-=(\alpha_1-\alpha_2)e^{i\omega t-i\phi_0}/\sqrt2$.
The collective mode is the common mode $\widetilde{\alpha}_1\equiv\alpha_+=(\alpha_1+\alpha_2)/\sqrt2$, with $n_{1,\text{ss}}=\gamma_{\uparrow}/\gamma_{\downarrow}=2r_{\text{LC}}^2$.
Since the synchronization measure $\Ss_c=1/\langle q_-^2+p_-^2\rangle$ involves only the relative mode, the entire synchronization dynamics of two oscillators is carried by this single nearly Gaussian mode.
The general rates reduce to those of the main text,
\begin{equation}
\lambda_{0,2}=2\kappa-\gamma_{\uparrow}\equiv\lambda_0,
\qquad
q_{0,2}=\kappa+\frac{\gamma_{\uparrow}}{2}\equiv q_0,
\label{Eq:Unified-N2-rates}
\end{equation}
and Eq.~\eqref{Eq:Unified-XY-solution} gives the relative-mode moments $X(t)$, $Y(t)$, which determine the OU estimate of the synchronization measure, $\Ss_c^{\mathrm{OU}}(t)=1/[X(t)+Y(t)]\equiv1/D_c(t)$.
For the initial states of the main text, $\alpha_1(0)=\eta_1r_{\text{LC}}$ and $\alpha_2(0)=\eta_2r_{\text{LC}}e^{i\Delta\phi_0}$, the initial moments follow from Eq.~\eqref{Eq:Unified-initial-moments} with $\beta_-(0)=[\alpha_1(0)-\alpha_2(0)]e^{-i\phi_0}/\sqrt2$ and $\phi_0=\arg\alpha_+(0)$.

The crossing time is fixed by $\Ss_c^A(t_\times)=\Ss_c^B(t_\times)$, or equivalently $\Delta D_c(t_\times)=0$, where the error difference $\Delta D_c(t)\equiv D_c^A(t)-D_c^B(t)$ of Eq.~\eqref{Eq:Unified-F} reduces to
\begin{equation}
\Delta D_c(t)=\Bs(t)+e^{-\lambda_0t}\Hs(t),
\label{Eq:Unified-N2-F}
\end{equation}
with $\Bs\equiv\Bs_2$ and $\Hs\equiv\Hs_2$ of Eq.~\eqref{Eq:Unified-BH}: these are precisely the slowly decaying background and the memory term identified in the main text.
To estimate the crossing time, we expand $\ell(t)=\ln|\Hs(t)/\Bs(t)|$ in powers of the dimensionless time $\tau=\gamma_{\uparrow}t$ as $\ell=\ell_0+\ell_1\tau+\frac12\ell_2\tau^2+\cdots$; the crossing condition $\ell(t_\times)=\lambda_0t_\times$ then becomes a quadratic equation in $\tau$, whose solution gives
\begin{equation}
\gamma_{\uparrow}t_\times\simeq
\frac{(\lambda_0/\gamma_{\uparrow}-\ell_1)-\sqrt{(\lambda_0/\gamma_{\uparrow}-\ell_1)^2-2\ell_2\ell_0}}{\ell_2},
\label{Eq:Unified-N2-tx}
\end{equation}
with the expansion coefficients given explicitly by
\begin{equation}
\ell_0=\ln\left|\frac{\Hs(0)}{\Bs(0)}\right|,
\qquad
\ell_1=\frac{\dot{\Hs}(0)}{\Hs(0)}-\frac{\dot{\Bs}(0)}{\Bs(0)},
\qquad
\ell_2=\frac{\ddot{\Hs}(0)}{\Hs(0)}-\left[\frac{\dot{\Hs}(0)}{\Hs(0)}\right]^2
-\frac{\ddot{\Bs}(0)}{\Bs(0)}+\left[\frac{\dot{\Bs}(0)}{\Bs(0)}\right]^2,
\label{Eq:Unified-N2-ell}
\end{equation}
where the dot denotes the derivative with respect to the dimensionless time $\tau=\gamma_{\uparrow}t$.
Among the two roots, the physically relevant one is selected by continuity with the linear-order solution in the limit $\ell_2\to0$; the second root is not continuously connected to the linear-order crossing and, for the parameters considered here, lies outside the time window in which the local expansion of $\ell(t)$ is accurate.
For the parameters of the main text, $\ell_0=4.63$, $\ell_1=-1.90$, and $\ell_2=2.06$, which gives $\gamma_{\uparrow}t_{\times}\simeq0.55$, in good agreement with the exact master-equation value $\gamma_{\uparrow}t_{\times}\simeq0.56$ reported in the main text.

The same numerical comparison verifies the Gaussian closure underlying the unified solution.
The exact relative-mode Wigner functions closely follow the contours of the Gaussian reference state that shares the same first and second moments [Fig.~\ref{Fig:N2RelatMW}(a)--(j)], and the non-Gaussianity quantified in the End Matter of the main text remains small at all times.
Accordingly, the OU estimate $\Ss_c^{\mathrm{OU}}$ deviates from the exact synchronization measure by at most a few times $10^{-3}$ at all times [Fig.~\ref{Fig:N2RelatMW}(k)].

Moreover, the analytical solution doubles as a screening tool for the SMpE in the space of system and initial-state parameters.
For $N=2$, the general late-time form~\eqref{Eq:Unified-B-late} reduces to $\Delta D_c(t)\simeq\Bs(t)\simeq Q'[1]\,(c_B-c_A)\,e^{-\gamma_{\uparrow}t}$, with $Q\equiv Q_2$, so the approach is from below whenever $Q'[1](c_B-c_A)<0$.
For the parameters of the main text, $Q'[1]\simeq-0.028$, so the nontrivial requirement is $c_B>c_A$: the initially less synchronized state must start with its common mode closer to the limit cycle, experiencing the stronger nonlinear damping of the relative mode.
The crossing time of the screened candidates is then estimated by Eq.~\eqref{Eq:Unified-N2-tx}.

\blueit{The $N>2$ case.--}
For a genuine network, all $N-1$ transverse modes contribute to the synchronization error, and within the OU description the many-body measure of Eq.~\eqref{Eq:many-body-sync} reads
\begin{equation}
D_\perp(t)=\Ss_\perp^{-1}(t)=\frac{1}{N-1}\sum_{\mu=2}^{N}[X_\mu(t)+Y_\mu(t)],
\label{Eq:Unified-Dperp}
\end{equation}
so that the error difference $\Delta D_\perp(t)\equiv D_\perp^A(t)-D_\perp^B(t)$ is the mode sum of Eq.~\eqref{Eq:Unified-F}.
Each mode relaxes at its own rate $\lambda_{0,\mu}=\lambda_\mu-\gamma_{\uparrow}$: the algebraic connectivity $\lambda_2$ sets the slowest transverse tail, while larger eigenvalues provide faster channels that can carry an initially large error away.
Consequently, the SMpE in a network is controlled by the spectral distribution of the Laplacian rather than by a single coupling constant.

This mode resolution becomes explicit when the two preparations start with the same collective-mode intensity, $n_1^A(0)=n_1^B(0)$, so that their collective backgrounds coincide at all times,
\begin{equation}
z_A(t)=z_B(t)\equiv z(t)
\qquad (c_A=c_B\equiv c),
\label{Eq:Unified-common-z}
\end{equation}
and share the same initial transverse covariance, differing only in their coherent transverse displacements, $\Delta\beta_\mu(0)=\beta_\mu^A(0)-\beta_\mu^B(0)$.
The background terms then cancel exactly, $\Bs_\mu=0$, and Eq.~\eqref{Eq:Unified-F} reduces to
\begin{equation}
\Delta D_\perp(t)=\frac{1}{N-1}\sum_{\mu=2}^{N}e^{-\lambda_{0,\mu}t}\bigl[\Delta X_\mu(0)R^3(t)+\Delta Y_\mu(0)R(t)\bigr].
\label{Eq:Unified-delta-Dperp}
\end{equation}
In this case, the late-time estimate~\eqref{Eq:Unified-B-late} becomes inconclusive, since both of its sides vanish identically; the existence of a crossing and the sign of the late-time approach are instead decided entirely by the transients, with the slowest mode $\mu=2$ dominating at late times.
For purely radial displacements, $\Delta Y_\mu(0)=0$, the common factor $e^{\gamma_{\uparrow}t}R^3(t)$ drops out of the crossing condition, leaving the spectral form
\begin{equation}
\sum_{\mu=2}^{N}\Delta X_\mu(0)\,e^{-\lambda_\mu t_\times}=0.
\label{Eq:Unified-network-crossing}
\end{equation}
Thus, if the difference is dominated by a fast mode $f$ and a slow mode $s$, with $\Delta X_f(0)=C_f>0$, $\Delta X_s(0)=-C_s<0$, and $\lambda_f>\lambda_s$, the crossing time is simply
\begin{equation}
t_\times=\frac{\ln(C_f/C_s)}{\lambda_f-\lambda_s}.
\label{Eq:Unified-two-mode-crossing}
\end{equation}
For a network such as the $N=10$ example considered in the main text, Eq.~\eqref{Eq:Unified-network-crossing} is evaluated as a sum over the nine transverse eigenmodes, while Eq.~\eqref{Eq:Unified-two-mode-crossing} gives a useful dominant-mode estimate.
Physically, the synchronization Mpemba effect results from an initially larger error being preferentially loaded into faster transverse Laplacian modes, whose rapid decay allows the apparently less synchronized state to overtake its competitor.

\blueit{Summary.--}
The unified solution~\eqref{Eq:Unified-XY-solution} reveals a common theme behind the SMpE at any system size: the initially less synchronized state overtakes its competitor because its excess error is preferentially loaded into faster-decaying error channels.
The nature of these channels, however, differs between the two cases.
For $N=2$, only a single transverse mode exists, and the competition takes place within it: the fast channel is the transient $e^{-\lambda_{0}t}\Hs(t)$, which carries the initial displacement and fluctuation mismatch at the rate $\lambda_0=2\kappa-\gamma_{\uparrow}$, while the slow channel is the background $\Bs(t)$, generated by the difference of the collective backgrounds and decaying on the gain timescale, $\Bs(t)\simeq Q'[1]\,(c_B-c_A)\,e^{-\gamma_{\uparrow}t}$.
For $N>2$, the transverse subspace provides a genuine spectral mechanism: the modes relax independently at the rates $\lambda_{0,\mu}=\lambda_\mu-\gamma_{\uparrow}$, and an initially larger error can be carried away simply by being concentrated in the faster modes, even when the two preparations share the same collective background, so that $\Bs_\mu=0$ identically [cf.~Eq.~\eqref{Eq:Unified-delta-Dperp}].
In this sense, the two-oscillator SMpE is a background--transient competition within a single mode, whereas the network SMpE is an intermode competition controlled by the Laplacian spectrum, whose simplest quantitative expression is Eq.~\eqref{Eq:Unified-two-mode-crossing}.

\bibliography{filterRef}